\documentclass[final,5p,times,twocolumn]{elsarticle}

\usepackage{amsthm}
\usepackage{dblfloatfix}
\usepackage{euscript}   
\usepackage{graphicx}
\usepackage{dcolumn}
\usepackage{bm}
\usepackage{amsmath}
\usepackage{amssymb}

\usepackage{mathptmx}
\usepackage{textcomp}
\usepackage{tabularx} 
\usepackage{times}
\usepackage{titlesec}
\usepackage{wrapfig}
\usepackage{caption}
\usepackage{subcaption}
\usepackage{adjustbox}
\usepackage{listings}
\usepackage{silence}
\usepackage{wasysym}
\usepackage{titlesec}
\usepackage{graphicx}
\usepackage{colortbl}
\usepackage[table,xcdraw]{xcolor}
\usepackage{booktabs}

\usepackage{graphicx}
\usepackage[utf8]{inputenc}
\usepackage[T1]{fontenc}
\usepackage{etoolbox}
\usepackage{csquotes}
\usepackage{xcolor}

\usepackage{hyperref}
\hypersetup{
    colorlinks=true,
    linkcolor=green,
    filecolor=magenta,      
    urlcolor=cyan}

\usepackage{siunitx}

\usepackage{diagbox}
\begin{document}
\title{Electronic and thermal properties of the phase-change memory material, Ge$_2$Sb$_2$Te$_5$, and results from spatially resolved transport calculations.}

\author[OU]{K. Nepal}
\ead{kn478619@ohio.edu}
\affiliation[OU]{organization={Department of Physics and Astronomy, Nanoscale and Quantum Phenomena Institute (NQPI)},
            addressline={Ohio University}, 
            city={Athens},
            postcode={45701}, 
            state={OH},
            country={USA}}

\author[OU]{A. Gautam}

\author[OU]{R. Hussein}

\author[Tampere]{K. Konstantinou}
\affiliation[Tampere]{organization={Department of Mechanical and Materials Engineering},
            addressline={University of Turku}, 
            city={Turku},
            postcode={FI-20500}, 
            country={Finland}}
            

\author[SRE]{S. R. Elliott}
\affiliation[SRE]{organization={Physical and Theoretical Chemistry Laboratory, Department of Chemistry},
            addressline={University of Oxford}, 
            city={Oxford},
            postcode={OX1 3QZ}, 
            country={UK}}

\author[LANL]{C. Ugwumadu}
\ead{cugwumadu@lanl.gov}
\affiliation[LANL]{organization = Physics of Condensed Matter and Complex Systems (T-4) Group, addressline = { Los Alamos National Laboratory}, city = { Los Alamos}, postcode = {87545}, state ={NM}, country={USA}}

\author[OU]{D. A. Drabold}
\ead{drabold@ohio.edu}

\date{July 2025}

\begin{abstract}
We report new insights into the electronic, structural, and transport (heat and charge) properties of the phase-change memory material Ge$_2$Sb$_2$Te$_5$. Using realistic structural models of Konstantinou \textit{et al.} [Nat. Commun. 10, 3065 (2019)], we analyze the topology, electronic states, and lattice dynamics with density functional methods, including hybrid-functional calculations and machine-learned interatomic potentials. The Kohn–Sham orbitals near the Fermi level display a strong electron–phonon coupling, and exhibit large energy fluctuations at room temperature. The conduction tail states exhibit larger phonon-induced fluctuations than the valence tail states. To resolve transport at the atomic scale, we employ space-projected electronic conductivity and site-projected thermal conductivity methods.  Local analysis of heat transport highlights the role of filamentary networks dominated by Te, with Sb and Ge making progressively smaller contributions.
\end{abstract}

\maketitle
\section{Introduction}
Germanium antimony telluride (Ge$_2$Sb$_2$Te$_5$), henceforth referred to as GST, is fascinating and vastly studied for the science of its rapid transformation between amorphous and crystalline states and associated electrical conductivity contrast between the two states, which enables its application for practical computer memory devices. 

Pioneering computer simulations of Hegedus and Elliott \cite{hegedus_and_Elliot2008} revealed that it is possible to study the nature of the phase change quite directly from accurate molecular dynamics (MD) simulations.  A vast amount of insight has accrued from this beginning \cite{lencer2008map,Luckas2013,Zhou2014, Mukhopadhyay2016, Marc2017, Konstantinou2019, Le2020, Mocanu2020, Konstantinou2022}. 

In this paper, we utilize realistic computer models of Konstantinou, Mocanu, Lee, and Elliott  (KMLE) \cite{Konstantinou2019} to explore electronic structure and dynamics using a hybrid functional, and we compute quantities of technological interest, including spatially local estimates of charge and heat transport in the materials. We include three Appendices which offer a short, self-contained description of the transport methods. Details on the transport calculations may be found in the original papers \cite{SPC1, SPC2, Gautam2024-SPTC-first-paper, Ugwumadu2025-SPTC-second-paper}.

The principal conclusions of this work are that (i) there are large thermally induced fluctuations of Kohn-Sham states near the Fermi level, highlighting the importance of electron–phonon coupling in electronic conductivity, (ii) electronic conductivity is spatially heterogeneous and often involves Sb-vacancy environments, and (iii) site-projected thermal conductivity analysis shows that Te- and Sb-rich regions preferentially contribute to heat transport. Together, these findings establish that electronic and thermal transport in GST are spatially inhomogeneous, significantly influenced by phonons, defect-sensitive, and governed by a small subset of structurally distinct motifs - insights that may provide information for tailoring phase-change memory materials at the atomic scale.

The remainder of the paper is organized as follows. Section \ref{sec:methods} describes the computational methodology, including model construction (relaxation), electronic-structure calculations, and vibrational analysis. We compute estimates of spatially resolved transport: Space-projected conductivity (SPC) and site-projected thermal conductivity (SPTC). We provide concise descriptions of these methods in \ref{app:spc} and \ref{app:sptc}, respectively. Section \ref{sec:results} presents the structural changes of the KMLE models under HSE06 relaxation, associated electronic structure, and defects.  We investigate thermally driven electronic fluctuations near the Fermi level, and report spatially resolved electronic and thermal transport. Section \ref{sec:conclusion} summarizes our findings and discusses broader implications for phase-change memory design. In the representation of atoms in the figures in this work, Ge, Sb, and Te are colored teal, purple, and brown, respectively.

\section{Methodology}\label{sec:methods}

\subsection{GST Structural Models}
Computations described in this paper used the \textit{ab initio} plane wave code Vienna Ab initio Simulation Package (VASP) \cite{VASP}. Twelve cubic supercells containing 315 atoms with chemical composition germanium (Ge), antimony (Sb), and tellurium (Te) in a 2:2:5 ratio due to KMLE \cite{Konstantinou2019} were considered in this work. These models are referred to as M1, M2, ..., and M12 for the rest of the paper. These GST models were initially generated through classical molecular dynamics simulations, implemented in LAMMPS (Large-scale Atomic/Molecular Massively Parallel Simulator) \cite{lammps} employing a machine-learned Gaussian approximation potential (GAP) developed by Mocanu \textit{et. al.} \cite{Mocanu2018}. A 900-atom model of GST also from KMLE is considered. 


\subsection{Topological and Electronic Structure}

To accurately compute the electronic structure and gap states near the Fermi level ($\epsilon_f$), we carried out relaxations and electronic structure calculations using the hybrid functional of Heyd, Scuseria, and Ernzerhof (HSE06) \cite{HSE1}.  The thermal MD performed in this work also used HSE06. Periodic boundary conditions were used in all calculations. A plane-wave basis set with a kinetic energy cut-off of 240 eV for geometrical relaxation was used. A kinetic energy cut-off of 520 eV was implemented for electronic structure calculation. The Brillouin zone of the supercell models was sampled at the $\Gamma$ point.

\subsubsection{Spatially local estimates of transport}

Electronic and thermal transport are determined by the network topology, chemical order, electronic, and vibrational properties of materials. For a disordered system, no two sites are identical, and it is unknown how much variation exists from site to site. For purposes of engineering materials with preferred transport properties, atomic-level insights may be helpful.

For both the thermal and electronic case we have made two assumptions: (1) the validity of a Kubo formula [either the Kubo-Greenwood Formula (KGF) or the Green-Kubo Formula (GKF) \cite{Kubo,Green1954-Green-Kubo-relation}], and assumptions implicit to these  (thermal equilibrium, linear response, and an expression for the heat or electron current) and (2) a spatial decomposition of the Kubo formula to provide local information about either form of transport. Our innovation here has involved only the second point and the formulation and interpretation of the electronic and thermal conductivity are discussed in \ref{app:spc} and \ref{app:sptc}, respectively.

\subsection{Vibrational Properties}
The vibrational signatures of the GST models were examined by computing the vibrational density of states (VDOS) and vibrational inverse participation ratio (VIPR) within the harmonic approximation (HO). The dynamic matrix (DM) was constructed from finite differences from the usual definition:

\begin{equation}\label{eq:DM_FC}
    D^{\alpha \beta}_{i j} = \frac{1}{\sqrt{m_i m_j}} \frac{\partial^2 E}{\partial u^{\alpha}_i \partial u^{\beta}_j}
\end{equation}

Here,  $u_i^{\alpha}$ is the small displacement of $i^{th}$ atom along cartesian ($\alpha$) direction, $m_i$ is the mass of the $i^{th}$ atom, and $E$ is the potential energy. The eigenvalue problem for the classical normal modes at the center (\textbf{k=0}) of the phonon Brillouin zone is \cite{Gautam2024-SPTC-first-paper}: 

\begin{equation}
        \omega^2_m \  e_i^{\alpha,m} = \sum_{\beta j} D^{\alpha \beta }_{i j} e_j^{\beta,m}
\end{equation}
where, $\omega_m$ and $e_i^{\alpha}$ are the vibrational frequency of the mode $m$ and sets of displacement directions for each atom in the mode $m$ at atom $i$ along the $\alpha$ direction. The Vibrational Inverse Participation Ratio (VIPR) ($g$) corresponding to the vibrational modes $m$ are: 
\begin{equation}
    g(m) = \frac{\sum_{i, \alpha} |e_{i}^{\alpha,m}|^4}{\big(\sum_{i, \alpha} |e_{i}^{\alpha,m}|^2\big)^2 }
\end{equation}
VIPR is a measure of localization, and varies between 1/N (extended vibrational mode) and 1 (ideally localized vibrational mode). To compute the DM, defined in Equation \ref{eq:DM_FC}, each atom in GST was displaced 0.05 \AA{} in x-, y-, and z- directions, from its energy-minimized configuration. The resulting modes were also employed to calculate the thermal conductivity and site-projected thermal conductivity.

\section{Results}\label{sec:results}
\subsection{Structural Relaxation}
\begin{figure*}[!t]
 \centering
	\includegraphics[width=.8\linewidth]{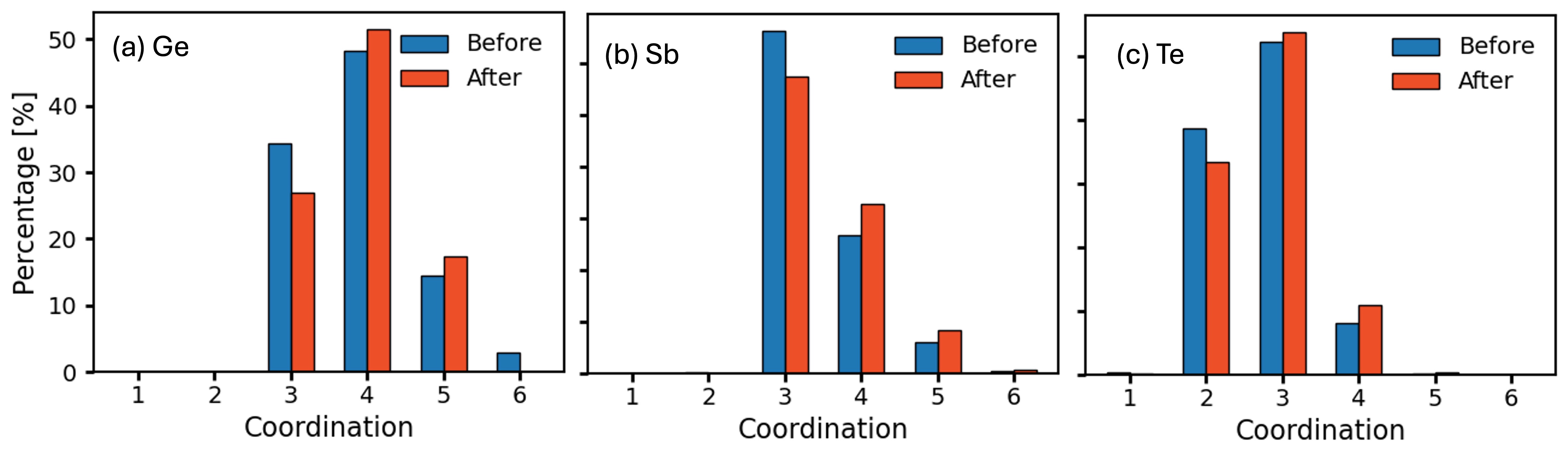}
	\caption{ Coordination analysis for the local environment of (a) germanium, (b) antimony, and (c) tellurium, respectively, averaged over twelve GST models from KMLE after HSE06 relaxation (Red color). Coordination number was calculated by using a cut-off bond distance of 3.2 \AA{} for each atom species. The blue histogram shows the average coordination calculated from before HSE06 relaxation.}
 \label{coordination}
\end{figure*}

In this section, we report the change in coordination in the GST models due to relaxation with HSE06. A radial cut-off distance of 3.2 \AA{} was used to define coordination. Figure \ref{coordination} illustrates the average coordination of atomic species before and after HSE06 relaxation in red and blue histograms, respectively. Before relaxation, it is observed that Ge prefers three- and four-fold coordination, while Sb favors three- and two-fold coordination, and Te favors three- and two-fold coordination, as shown by the red histograms in Figure \ref{coordination}. Following relaxation, the basic features remain unchanged; however, a reduction in three-fold coordination in Ge and Sb was observed with an increase in four- and five-fold coordination environments, which suggests formation of defective octahedral motifs. This is supported by the Ge-centered and Sb-centered angle distribution, which shows a pronounced increase around $\approx$ 170$^\circ$, confirming that the higher coordination arises from these distorted octahedral environments \cite{Lee2021}, see Figure  \textcolor{blue}{S2}. For Te, increased three- and four-fold coordination was observed with reduced two-fold coordination . The findings are shown respectively as blue and red histograms in Figure \ref{coordination}.

\subsection{Defects in GST}
The electronic structure of the models is sensitive to the choice of exchange–correlation functional and pseudopotential within density functional theory (DFT) \cite{Xu2009, Mukhopadhyay2016}. All models were relaxed and analyzed using the HSE06 functional. KMLE  \cite{Konstantinou2019, Konstantinou2022} used a similar hybrid-functional calculation to examine electronic structure. Here, we report the relatively minor changes in structure resulting from HSE06, and also carry out computationally demanding thermal MD simulations with HSE06. 

 The electronic structure nomenclature is defined by four categories: clean gap, shallow gap,  mid-gap, and multiple mid-gap. Of the twelve models, five have gap states. Of the five models with gap states, three have a shallow gap state near the conduction band edge, three models have one mid-gap state, and one model has two mid-gap states. Where electronic structure is concerned, the only difference between this work from KMLE involved M4, which showed no mid-gap state.  The band gap in an amorphous material is usually described as the Mott gap (the energy difference between the conduction and valence mobility edges). In this paper, we adopt the HUMO-LUMO energy difference as the effective gap. The positions of these defect states with respect to the lowest unoccupied molecular orbital (LUMO) are tabulated in Table \ref{tab:GapStates}. The calculated band gaps range from 0.42 - 0.69 eV with an average value of 0.59 eV,  in reasonable agreement with other simulations \cite{Konstantinou2019, Caravati2009} which report values between 0.5-0.8 eV and slightly lower than experimental values ($>$ 0.7 eV) \cite{Lee2005, Kato2005}.  The shallow defect level is 0.12-0.15 eV below the LUMO, which is slightly lower than the experimentally reported value of 0.18 eV \cite{Luckas2013}, and KMLE. 


\begin{figure*}[!t]
    \centering
	\includegraphics[width=.75\linewidth]{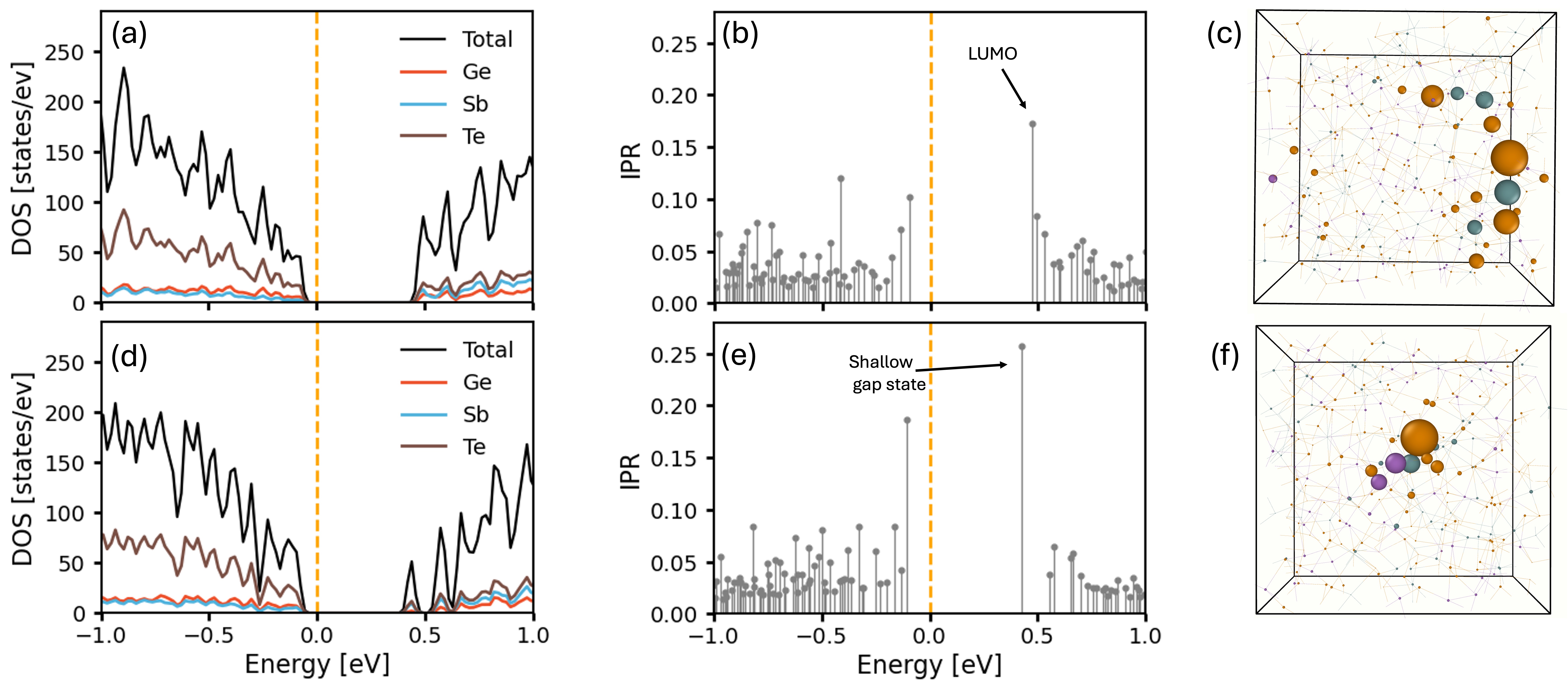}
	\caption{(a) The total and projected density of states near the band edges for M3. The projected density of states for atomic species reveals that Te dominates the band edges. (b) The IPR indicates the extent of electronic localization of states near the band edges. (c) Atom-projection of the states at the conduction band edge (LUMO), labeled in (b). The conduction tail state is predominantly due to Ge-Te and Ge-Ge, Te-Te forming a network illustrated by a large sphere. The volume of the sphere represents the atomic contribution to the conduction band tail state. (d)  The total and projected density of states near the band edges for M7 show a shallow-gap defect near the conduction band edge. (e) The shallow gap state is localized (IPR $\approx$ 0.25). (f) Atom-projection of the shallow gap state is predominantly due to a Te atom (the largest brown sphere in (f)), and Sb-Sb bonds.  Color code: teal - Ge, purple - Sb, and brown - Te.  }
 \label{fig:kfig_clean_shallow_gap}
\end{figure*}

\begin{figure*}[!t]
	\centering
    \includegraphics[width=.75\linewidth]{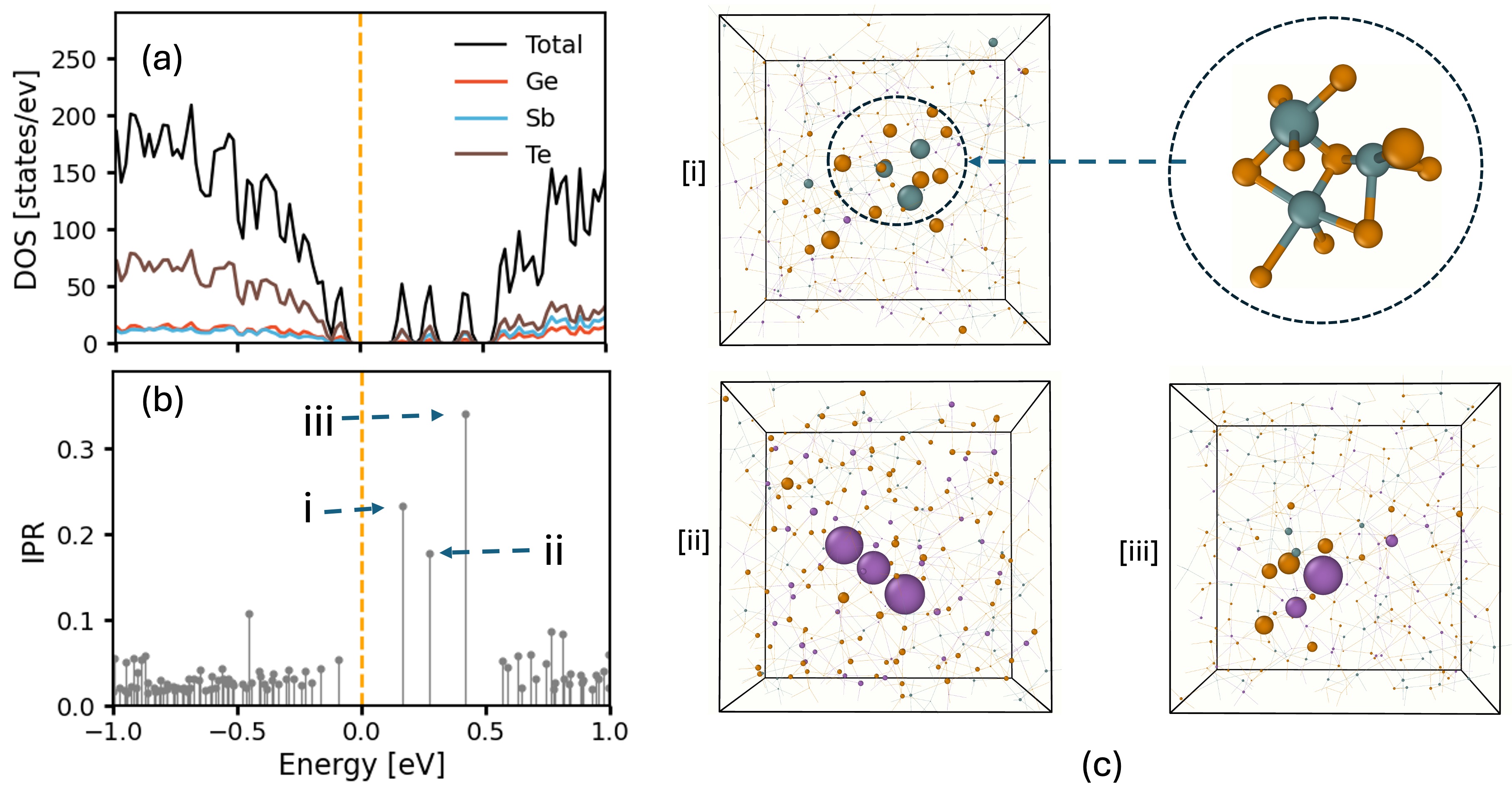}
	\caption{(a)  The total and projected density of states near the band edges for the GST model M8 shows multiple defects in the HOMO-LUMO gap. (b) Mid-gap gap states ("i-ii") and a shallow-gap defect ("iii")  are localized, indicated by high IPR values. (c)[i-iii] Atom-projection of the mid-gap states and a shallow-gap state. The mid-gap state "i" is predominantly due to a cluster of over-coordinated Ge atoms with Te atoms (cluster is highlighted). Second mid-gap state "ii" 0.11 eV above the mid-gap level is localized on a Sb-Sb-Sb trimer structure, while shallow-defect state "iii" 0.15 eV below the LUMO level is centered at a group of Sb and Te atoms, shown in (c) [ii-iii], respectively. The volume of the sphere in (c) indicates the weight of atomic contributions to the defect state. Color code: teal - Ge, purple - Sb, and brown - Te. }
 \label{kfig_three_gap}
\end{figure*}

\subsubsection{Clean gap GST model}
 Seven out-of-twelve GST models exhibit a clean gap. We show an example for M3 in Figure \ref{fig:kfig_clean_shallow_gap}a. The electronic inverse participation ratio (IPR)  gauges electronic localization, analogous to VIPR defined above\footnote{This is carried out rather crudely in this paper, using the atom-projected atomic orbitals from VASP. It would be preferable to apply a more sophisticated method like LOBSTER \cite{Nelson2020}}. The valence band tail is mostly derived from Te atoms, whereas all the atomic species contribute to the conduction band tail. The  IPR is indicated by gray vertical lines in Figure \ref{fig:kfig_clean_shallow_gap}b. States near the conduction band minimum and valence band maximum are significantly localized. The atomic projection of a conduction band tail state (LUMO), labeled "i" in Figure \ref{fig:kfig_clean_shallow_gap}b, shows a contribution from a chain of Ge and Te atoms forming a filament of length $\approx$ 9 atoms. Such filamentary motifs are consistent with prior reports that near-linear Te–Ge–Te units act as favorable sites for electron localization in the conduction band tail \cite{Konstantinou2022_}. This is illustrated in Figure \ref{fig:kfig_clean_shallow_gap}c, where the size of the sphere represents the weights of the atom to the LUMO.

\begin{table}[t!]
\centering
\scriptsize
\caption{{band gaps, defect counts, and defect energy positions for 12 GST models (HSE06).}}
\footnotesize
\renewcommand{\arraystretch}{1.15}
\begin{tabularx}{\linewidth}{@{\extracolsep\fill}cccc@{}}
\toprule
\textbf{Model} & \textbf{band gap [eV]} & \textbf{Mid-gap\textsuperscript{a} [eV]} & \textbf{Shallow\textsuperscript{a} [eV]} \\
\midrule
M1  & 0.60  & — & — \\
M2  & 0.47  & 0.18 & — \\
M3  & 0.57  & — & — \\
M4  & 0.55  & — & — \\
M5  & 0.54  & 0.31 & — \\
M6  & 0.64  & — & — \\
M7  & 0.66  & — & 0.13 \\
M8  & 0.66  & 0.30; \;0.40 & 0.15 \\
M9  & 0.69  & 0.31 & 0.12 \\
M10 & 0.62  & — & — \\
M11 & 0.42  & — & — \\
M12 & 0.62  & — & — \\
\bottomrule
\end{tabularx}
\label{tab:GapStates}
\vspace{0.25cm}
\begin{minipage}{\linewidth}
\footnotesize \textsuperscript{a}Energies are reported as offsets \emph{below the LUMO} (i.e., deeper into the gap toward the Fermi level). Mean band gap across models: \(0.59 \pm 0.08\) eV.
\end{minipage}
\end{table}

\subsubsection{Shallow-gap defect GST model}
Three GST models (M7, M8, and M9) exhibited a shallow-donor level near the conduction edge. The total and projected density of states of the GST M7 is shown in Figure \ref{fig:kfig_clean_shallow_gap}d, showing a shallow donor state at $\approx$ 0.13 eV below the LUMO. The level is localized, as indicated by the high IPR value in Figure \ref{fig:kfig_clean_shallow_gap}e. Projection of the shallow-gap defect state onto atomic species reveals localization primarily on a cluster of atoms formed by Te atoms and neighboring Sb-Sb and Te-Te homopolar bonds, as illustrated in Figure \ref{fig:kfig_clean_shallow_gap}f. The size of the sphere in Figure \ref{fig:kfig_clean_shallow_gap}f corresponds to the weight of the contribution by the atom to the shallow gap. Similar shallow defects have previously been attributed to homopolar bonding \cite{Konstantinou2022}.

\subsubsection{Mid-gap defect GST model}

Three GST models exhibit a mid-gap defect deep in the band gap. The total density of states for model  M5 is shown in Figure \textcolor{blue}{S1}a. Note the mid-gap state, labeled as "i".  The projection of the defect state into atomic contributions revealed that over-coordinated Ge atoms are responsible for these deep defects: see Figure \textcolor{blue}{S1}b, where the size of the sphere shows the weight of the atom species to the mid-gap state. The structural motif of these Ge atoms is the local crystal-like bonding environment in the amorphous network (see Figure \textcolor{blue}{S1}c), where the Ge atom is responsible for the state in a rock-salt environment. This observation has been previously reported by Reference \cite{Konstantinou2019}. Reference \cite{Konstantinou2019} also demonstrated that these mid-gap defects are capable of creating deep electron traps within the band gap.

    

\subsubsection{Multiple mid-gap defect GST model}

\begin{figure*}[!t]
\centering
	\includegraphics[width=0.8\textwidth]{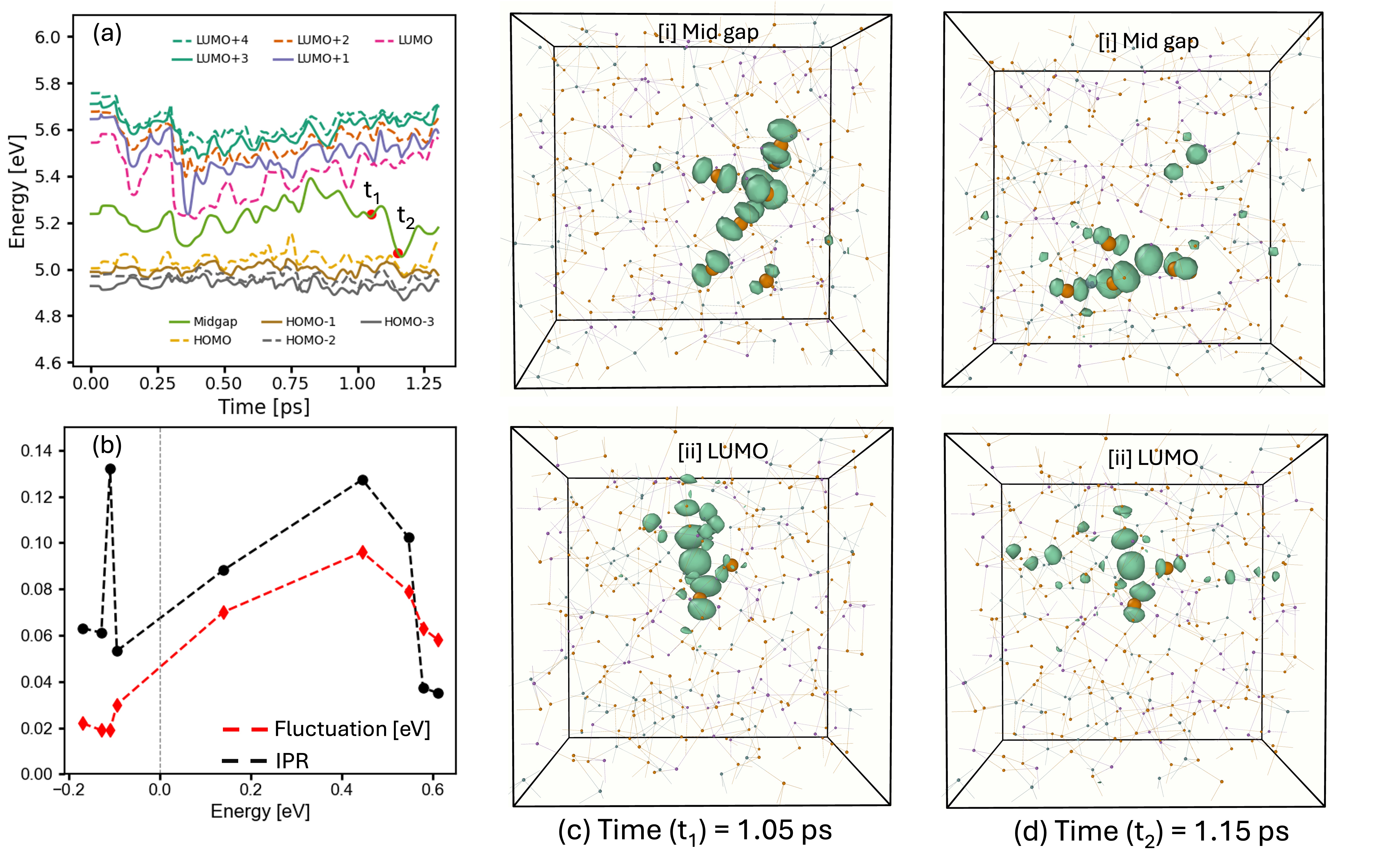}
	\caption{All plots correspond to M5 thermally equilibrated at 300K using a \text{Nos\'e}-Hoover thermostat with a time step of 1.5 fs. (a) Fluctuation in Kohn-Sham orbitals near the Fermi level, depicting fluctuations in LUMO (pink), mid gap (green), and HOMO (dashed-yellow). See legends for fluctuations in different orbitals. (b) Energy-resolved orbital properties for selected states in GST model M5. Black markers and lines show the inverse participation ratio (IPR), indicating the degree of localization of each state. The red dashed line with diamond markers represents the thermal fluctuation of the corresponding orbital energies. (c-d) Snapshots of mid-gap and LUMO energy level as an isosurface plot at times "t$_1$" and "t$_2$", respectively, labeled in (a).}
    
    \label{fig:kfig_300K_Transport}
\end{figure*}

The electronic density of states for M8 is shown in Figure \ref{kfig_three_gap}a. The EDOS shows two mid-gap defects at 0.4 and 0.3 eV below the LUMO, with additional shallow defect states at 0.15 eV below the LUMO. These defect states are localized as illustrated by the high vertical line for the IPR in Figure \ref{kfig_three_gap}b. To study the structural motif of these gap defects, each state, labeled as "i-iii", is projected onto atomic species, as shown in Figure \ref{kfig_three_gap}c[i-iii], respectively, where the size of the sphere indicates the weight of the atoms contributing to the defect states. The projection shows that a cluster involving over-coordinated Ge atoms contributes to the mid-gap state 0.4 eV below LUMO. The local structure of the cluster is highlighted as a magnified image in Figure \ref{kfig_three_gap}c [i]. Two five-fold and one four-fold Ge atoms bonded with Te atoms contribute to the mid-gap defect. The projection of mid-gap state, labeled "ii", onto atomic species, shows that the state is localized in a Sb$-$Sb$-$Sb bond network, see Figure \ref{kfig_three_gap}c[ii], while a group of Sb and Te atoms contributes the most to shallow defect state "iii", see Figure \ref{kfig_three_gap}c[iii].

\subsection{Electron-phonon coupling and conductivity}\label{}

The electron-phonon coupling is large for localized states \cite{AttaFynn2004, Prasai2016}, and this has implications to charge transport. This is related to Thomas' concept of "phonon-induced delocalization. \cite{Overhof1989,Thomas2000}. Thomas' view is justified by time-dependent DFT calculations in amorphous silicon  (a-Si). \cite{Li2003, DAD_119}. It seems likely that the effect is relevant in GST as well. 

Within an adiabatic approximation \cite{Abtew2007, Drabold1999}, we discuss thermally induced fluctuations in the states around the Fermi level, considering the GST model M5 with a single mid-gap defect. Model M5 is thermally equilibrated at room temperature (300K) for 1.3 ps using the HSE06 functional with \text{Nos\'e}–Hoover thermostat \cite{Nos1984,Hoover1985} at a time step of 1.5 fs. Refer to Figure \textcolor{blue}{S3}a for the average root-mean-square deviation (RMSD) of each atomic species. The fluctuation of the band tail states with time is provided in Figure \ref{fig:kfig_300K_Transport}a, showing dramatic fluctuation of the LUMO and higher energy levels. Large thermally driven fluctuations in the value of the HOMO-LUMO gap varying between 0.18 eV and 0.55 eV are observed. The position of the \textcolor{black}{wandering mid-gap state (formally the LUMO level) fluctuates of order $\approx$ $\pm$ 0.3 eV, as illustrated by the dashed pink plot in Figure \ref{fig:kfig_300K_Transport}a. The mid-gap states vary between close proximity to the HOMO and LUMO+1. States above $\epsilon_f$ fluctuate more than valence tail states, a situation also seen in a-Si \cite{Drabold1991}. \textcolor{black}{For the first five states above $\epsilon_f$ , the RMS fluctuation is correlated with localization -- higher localization, more fluctuation. The pattern is less clear for states at or below the Fermi level, which uniformly show less fluctuation than low-lying conduction levels (see Figure \ref{fig:kfig_300K_Transport}b). The reason for this is uncertain, though probably connected to the fact that valence states contribute Hellman-Feynman forces to the ions, whereas the states above $\epsilon_f$ do not. }}

These calculations suggest that the band tails of GST, especially the conduction tail, might exhibit a significant temperature dependence \cite{Drabold1991}. Perhaps this could be experimentally probed using total photoelectron yield spectroscopy (for the case of amorphous silicon, see the work of Aljishi \textit{et al.} \cite{Aljishi1990}). A T-dependent conductivity can be estimated using a thermally averaged version of the $N^2$ method \cite{N2}, briefly reiterated in \ref{app:N2}.

\begin{figure}[!t]
	\includegraphics[width=\linewidth]{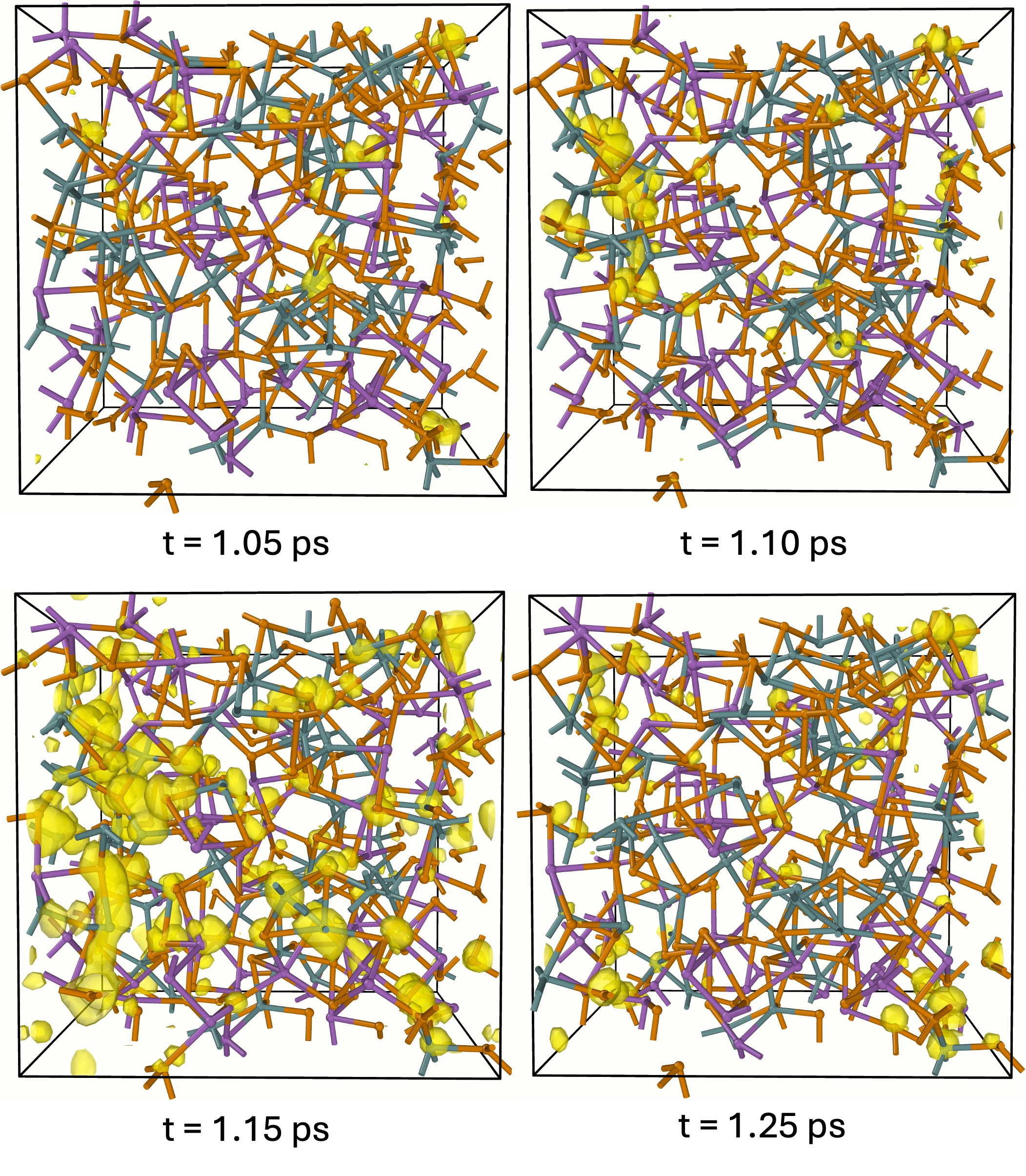}
	\caption{Thermally driven electronic conductivity fluctuations for M5 at 300K. The SPC  isosurface plot reveals fluctuations in conduction-active regions in the model at different times.  Color code: teal -Ge, purple - Sb, and brown - Te. Sb is shown as large spheres to identify antimony-rich and poor sites.}
    
    \label{kfig_spc_thermal}
\end{figure}

\begin{table*}[!b]
    \centering
    \scriptsize
    \caption{Atomic species resolved statistics for all 12 GST models, showing the minimum ($\zeta_{\mathrm{min}}$), mean ($\langle \zeta\rangle_{\mathrm{i}}$, with standard deviation, $\sigma_{\zeta}$), maximum ($\zeta_{\mathrm{max}}$) values, as well as per-element thermal conductivity ($\kappa_i$). The "Avg." row shows the average value per column.}
    \renewcommand{\arraystretch}{1.15}

    \begin{tabularx}{\linewidth}{@{\extracolsep\fill}c|cccc|cccc|cccc@{}}
          & \multicolumn{4}{c|}{ Ge [$\times$ 10$^{-3}$ W/m$\cdot$K] }  & \multicolumn{4}{c|}{ Sb [$\times$ 10$^{-3}$ W/m$\cdot$K]  } & \multicolumn{4}{c}{ Te [$\times$ 10$^{-3}$ W/m$\cdot$K] } \\

    Model & $\zeta_{\mathrm{min}}$ & $\langle \zeta\rangle_{\mathrm{Ge}}$ ($\sigma_{\zeta}$) & $\zeta_{\mathrm{max}}$ & $\kappa_{\mathrm{Ge}}$ &
    
    $\zeta_{\mathrm{min}}$ & $\langle \zeta\rangle_{\mathrm{Sb}}$ ($\sigma_{\zeta}$) & $\zeta_{\mathrm{max}}$  & $\kappa_{\mathrm{Sb}}$ & 
    
    $\zeta_{\mathrm{min}}$ & $\langle \zeta\rangle_{\mathrm{Te}}$ ($\sigma_{\zeta}$) & $\zeta_{\mathrm{max}}$ & $\kappa_{\mathrm{Te}}$ \\
    
    \hline
M1 &  0.21 & 0.41 (0.11) & 0.80 & 28.9 &  0.22 & 0.46 (0.11) & 0.72 & 32.5 &  0.26 & 0.52 (0.12) & 0.87 & 91.4 \\
M2 &  0.24 & 0.42 (0.09) & 0.65 & 29.3 &  0.23 & 0.47 (0.12) & 0.96 & 32.7 &  0.25 & 0.50 (0.10) & 0.85 & 87.0 \\
M3 &  0.19 & 0.40 (0.12) & 0.69 & 28.2 &  0.26 & 0.47 (0.11) & 0.88 & 33.0 &  0.30 & 0.51 (0.10) & 0.84 & 89.4 \\
M4 &  0.21 & 0.41 (0.10) & 0.75 & 28.7 &  0.23 & 0.48 (0.12) & 0.82 & 33.7 &  0.23 & 0.51 (0.10) & 0.79 & 88.6 \\
M5 &  0.25 & 0.41 (0.09) & 0.65 & 29.0 &  0.25 & 0.49 (0.10) & 0.71 & 34.0 &  0.23 & 0.53 (0.11) & 0.86 & 92.5 \\
M6 &  0.22 & 0.42 (0.08) & 0.60 & 29.3 &  0.26 & 0.47 (0.09) & 0.67 & 32.9 &  0.29 & 0.50 (0.09) & 0.83 & 86.9 \\
M7 &  0.15 & 0.38 (0.09) & 0.65 & 26.3 &  0.23 & 0.45 (0.08) & 0.65 & 31.6 &  0.28 & 0.51 (0.10) & 0.84 & 90.0\\
M8 &  0.24 & 0.42 (0.10) & 0.63 & 29.6 &  0.30 & 0.50 (0.08) & 0.69 & 34.7 &  0.28 & 0.54 (0.12) & 0.95 & 95.0 \\
M9 &  0.24 & 0.41 (0.08) & 0.60 & 28.7 &  0.29 & 0.45 (0.10) & 0.65 & 31.2 &  0.16 & 0.51 (0.10) & 0.77 & 88.9 \\
M10 &  0.12 & 0.43 (0.21) & 1.30 & 30.4 &  0.04 & 0.48 (0.18) & 0.90 & 33.7 &  0.01 & 0.55 (0.24) & 2.13 & 96.5\\
M11 &  0.25 & 0.43 (0.10) & 0.72 & 30.1 &  0.24 & 0.47 (0.10) & 0.73 & 33.2 &  0.30 & 0.52 (0.10) & 0.80 & 91.8 \\
M12 &  0.15 & 0.42 (0.10) & 0.64 & 29.2 &  0.27 & 0.48 (0.10) & 0.80 & 33.5 &  0.28 & 0.53 (0.11) & 0.87 & 92.7 \\
\hline
Avg. & 0.21 & 0.41 (0.10) & 0.72 & 29.0 & 0.24 & 0.47 (0.11) & 0.77 & 33.1 & 0.24 & 0.52 (0.11) & 0.95 & 90.9 \\
    \hline
    \end{tabularx}
    \label{tab:sptcStats}
\end{table*}

We present snapshots of Kohn–Sham orbitals at selected time steps to illustrate the evolution of a mid-gap level and the LUMO. Two representative times, labeled t$_1$ and t$_2$, are marked by red dots in the mid-gap level plot in Figure \ref{fig:kfig_300K_Transport}(a). During this interval, the mid-gap level shows pronounced fluctuations compared to the relatively stable LUMO.

The corresponding orbital snapshots are shown in Figure \ref{fig:kfig_300K_Transport}c and \ref{fig:kfig_300K_Transport}d, taken 100 fs apart.  The eigenvectors, visualized as green isosurface plots, in the mid-gap level reveal two distinct chain-like networks within the system, whereas the LUMO undergoes only minor structural changes. Importantly, the localized Kohn–Sham orbitals (mid-gap and LUMO) exhibit large thermally induced fluctuations, leading to substantial variations in instantaneous charge density. We further observe that the eigenvectors of the mid-gap level arise from two separate networks centered on overcoordinated Ge atoms.

\textcolor{black}{The influence of electron–phonon interaction on electronic conductivity is examined via SPC analysis (see \ref{app:spc}), obtained at different snapshots along a thermal MD trajectory.  Six Born-Oppenheimer snapshots (spaced by 75 fs, sampled from the final 1.3 ps of the 300 K MD simulation) were selected for SPC computations. The electronic conductivity spans a wide range, from 10$^{-5}$ to 10$^{-1}$ S/cm. Although the Kubo–Greenwood formula predicts average conductivities comparable to experimental reports ($\approx$ 10$^{-3}$ S/cm \cite{Ma2018}), the large variability among snapshots reflects fluctuations in the (HOMO–LUMO) gap and in the localization of states near the Fermi level. The gap fluctuates significantly, ranging from 0.18 eV to 0.55 eV  (see Figure \textcolor{blue}{S3}b). The SPC for four of the snapshots is shown in Figure \ref{kfig_spc_thermal}, illustrating how conduction-active regions evolve with lattice vibrations. SPC depicts the fluctuation in electronic conductivity that evolves from a low-conducting state (at times 1.05 ps and 1.10 ps) to a comparatively higher conducting state (at 1.15 ps) and reverts to a poor-conducting state (at time 1.25 ps).   Refer to Figure \textcolor{blue}{S4} for the atomic species contribution to the SPC.}

\subsection{Thermal transport in GST}

Site-projected thermal conductivity evaluations were conducted at room temperature (300 K) for the GST models. The SPTC extracted for the GST models in Table \ref{tab:sptcStats} and mirrored in Figure \textcolor{blue}{S5}a (Ge), \textcolor{blue}{S5}b (Sb), and \textcolor{blue}{S5}c (Te), indicates systematic species ordering, $\langle \zeta\rangle_{\mathrm{Te}} > \langle \zeta\rangle_{\mathrm{Sb}} > \langle \zeta\rangle_{\mathrm{Ge}}$, across nearly all structures. From the “Avg.” row, the mean values are $0.52$ $\times$ 10$^{-3}$ W/m$\cdot$K (Te), $0.47$ $\times$ 10$^{-3}$ W/m$\cdot$K (Sb), and $0.41$ $\times$ 10$^{-3}$ W/m$\cdot$K (Ge); thus Te carries, on average, $\approx$ 11 $\times$ 10$^{-5}$ W/m$\cdot$K absolute more SPTC than Ge ($\approx\!27\%$ higher) and $\approx$ 5  $\times$ 10$^{-5}$ W/m$\cdot$K  more than Sb ($\approx\!11\%$ higher). The reported intra-species dispersions are comparable (mean $\sigma_\zeta \approx 1{-}1.1$ $\times$ 10$^{-4}$ W/m$\cdot$K ), giving coefficients of variation\footnote{Coefficient of variation is the ratio to standard deviation to the mean} clustered near $0.24$ (Ge), $0.23$ (Sb), and $0.21$ (Te)--typical for amorphous chalcogenides with heterogeneous local environments \cite{Aryana2021tuning,Moon2026}.

\begin{figure}[!t]
    \centering
    \includegraphics[width=.9\linewidth]{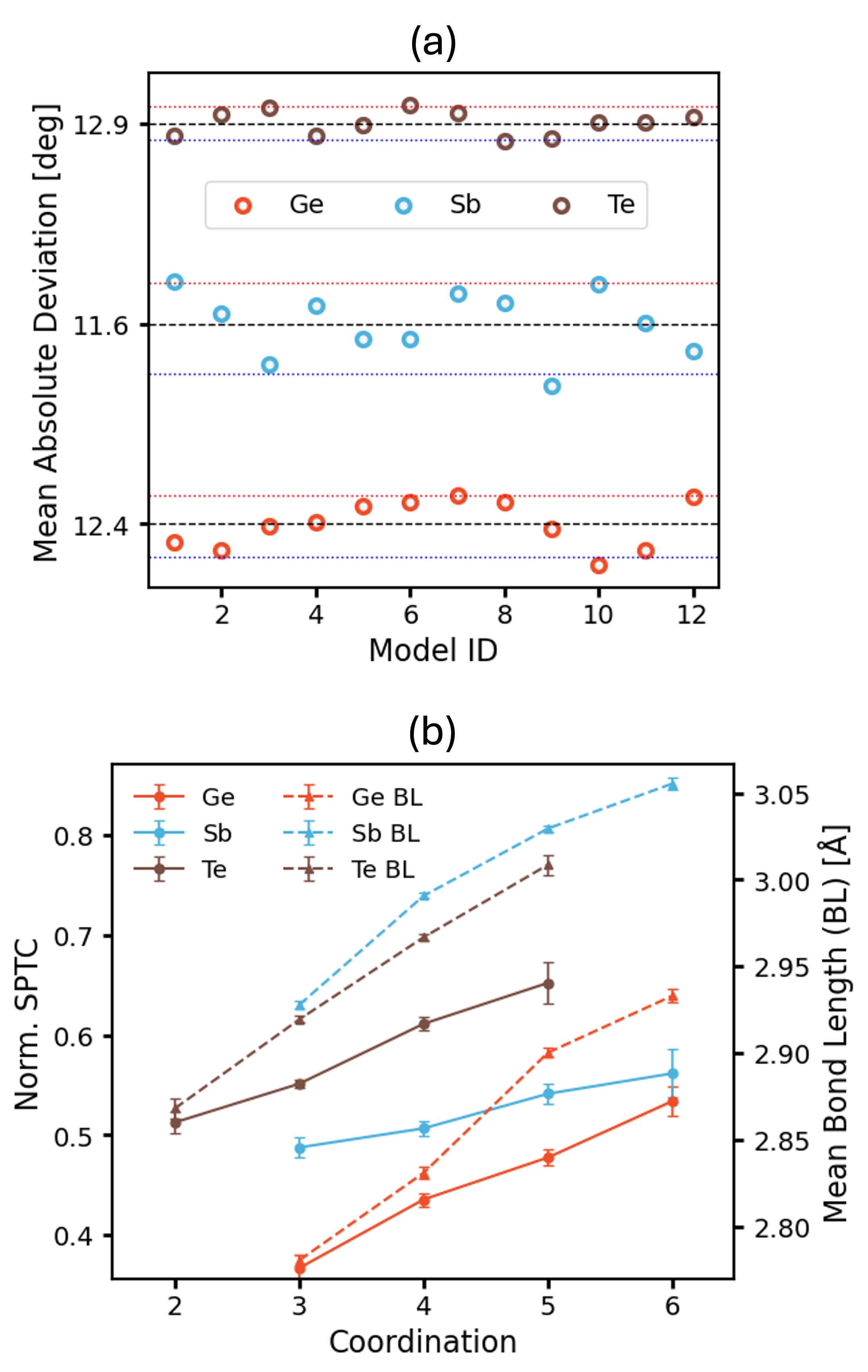}
    \caption{Thermal transport character in GST. (a) Mean absolute deviation (MAD) from rock-salt three-body angles (90$^\circ$ between two adjacent ions in the octahedron's equatorial plane) for each centered element. The blue and red dotted lines represent the 5$^\mathrm{th}$ (blue) and 95$^\mathrm{th}$ (red) percentiles of the distribution. (b) Correlation between (normalized) SPTC, bond length (BL), and coordination number per species averaged for models M1 to M12.}
    \label{fig:kfig_SPTC1}
\end{figure}

The extrema ($\zeta_{\mathrm{max}}$) further reinforce this hierarchy with the average of per-model $\kappa$ maxima being $0.95$ $\times$ 10$^{-3}$ W/m$\cdot$K for Te versus  $0.77$ $\times$ 10$^{-3}$ W/m$\cdot$K for Sb and $0.72$ $\times$ 10$^{-3}$ W/m$\cdot$K for Ge (also see Figure \textcolor{blue}{S5} a, b, and c). The element-averaged $\kappa$ mirror the same pattern: $\kappa_{\mathrm{Te}}\!\approx\!0.91$ W/m$\cdot$K, about three times $\kappa_{\mathrm{Sb}}$ ($\approx\!0.33$ W/m$\cdot$K) and $\kappa_{\mathrm{Ge}}$ ($\approx\!0.29$ W/m$\cdot$K). Of course, the Te carries more heat than the other species from the stoichiometry of the material.

 SPTC variance outliers highlight the structural sensitivity of amorphous GST. Model M10, in particular, shows the broadest intra-species spreads (Te: $\sigma_\zeta = 0.24$; Ge: $\sigma_\zeta = 0.21$; Sb: $\sigma_\zeta = 0.18$), reflecting a highly heterogeneous transport landscape. The angular metrics in Figure \ref{fig:kfig_SPTC1} (a) provide a structural explanation: larger deviations from the ideal GST rock-salt three-body angle \cite{Sun2006,Cinkaya2011} correlate with suppressed SPTC, and it is characteristic of angles in defective octahedral motifs \cite{Mocanu2020}. The black lines mark the mean angular deviations (MAD) of $11.6^\circ$, $12.4^\circ$, and $12.9^\circ$ for Sb, Ge, and Te, respectively (see Table \textcolor{blue}{S1}). The lower (blue) and upper (red) dashed lines indicate the 5$^\mathrm{th}$ and 95$^\mathrm{th}$ percentiles of the distributions, so atoms near the blue line are closer to the ideal rock-salt geometry, whereas those near the red line deviate most strongly.

A global comparison of the angle deviations shows that, although Te has the highest MAD of $\approx 13^\circ$, the spread around this mean is relatively narrow. This indicates a uniformly disordered Te environment, which has been reported to promote higher SPTC compared to more localized disorder, as seen in amorphous Si \cite{Ugwumadu2025-SPTC-second-paper} and in the contrast between vacancy and Frenkel defects in crystalline W \cite{W}. For Ge, the element-level conductivity is correlated with angular disorder: $\kappa_{\mathrm{Ge}}$ is lowest in M7, which exhibits the largest MAD, and highest in M10, which shows the smallest MAD. By contrast, the SPTC of Sb appears largely insensitive to angular deviations, consistent with its mean angle remaining close to the ideal rock-salt value.

This ordering is physically consistent with the established picture of heat transport in amorphous GST. At room temperature, thermal conduction is dominated by diffuson-like vibrational modes rather than well-defined propagating phonons; in the Allen–Feldman framework \cite{phonon1,phonon2}, the SPTC depends on both the vibrational density of states (VDOS) in the mid- to low-frequency window and $\Xi$. Figure~ \textcolor{blue}{S6} illustrates this clearly: Te and Sb contribute disproportionately in the low–mid frequency range where diffuson-like modes dominate amorphous transport. 

Within the SPTC formulation, diffuson-like modes contribute to site conductivities through the real-symmetric transport matrix $\Xi$ \cite{Ugwumadu2025-SPTC-second-paper}. In contrast, the VIPR curves (gray lines) associated with Ge peaks in the high-frequency range reveal strong vibrational localization (the locon regime), and such modes contribute little, or nothing, to heat transport \cite{Ugwumadu2025-SPTC-second-paper}. The same VDOS/VIPR pattern is consistently observed across the other eleven models (see Figure \textcolor{blue}{S7}). An outlier was noted for M9 (localized Ge-Ge stretch modes) in Reference \cite{animation}. 

We have shown that the range of the  "thermal matrix" $\Xi$ (see \ref{app:sptc}) is about $\approx$8~\AA{}. This quantifies the non-locality of heat transport in GST.  To estimate the site-wise contribution of the thermal conductivity requires information only within this range.

\begin{figure}[!t]
    \centering
    \includegraphics[width=.8\linewidth]{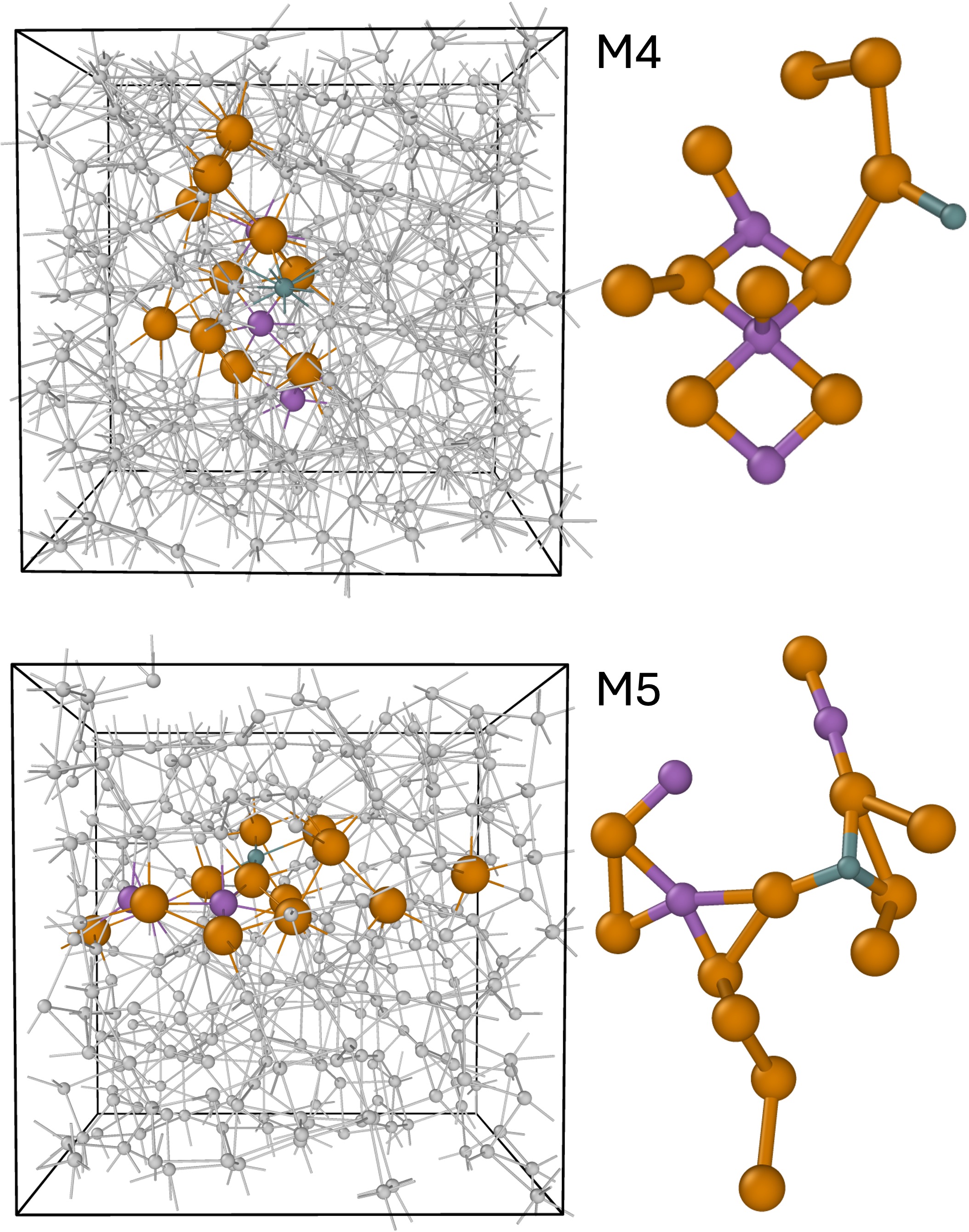}
    \caption{High SPTC ($\geq$ 70\% of maximum SPTC) filamentary structure in M4 and M5.  The atoms are color-coded: teal, purple, and brown for Ge, Sb, and Te, respectively.}
    \label{fig:kfig_SPTC2}
\end{figure}

This same argument for connected, percolative pathways is reinforced by Figures \ref{fig:kfig_SPTC1} (b) and \ref{fig:kfig_SPTC2}. In Figure \ref{fig:kfig_SPTC1} (b), SPTC is shown to increase with coordination number (CN) and, more weakly, with mean bond length (BL), with the strongest dependence for Te, intermediate for Sb, and weakest for Ge. These correlations demonstrate that higher coordination and favorable bonding distances provide the structural conditions most conducive to efficient phonon vibration and heat transfer. For per-model comparison, Figure \textcolor{blue}{S8} shows the percentage distribution of coordination numbers for the different elements. Consistent with this picture, the atoms responsible for the top-tier SPTC values (those $\geq 70\%$ of the per-model maximum) form extended filamentary networks rather than isolated clusters. As illustrated for models M4 and M5 in Figure \ref{fig:kfig_SPTC2}, these space-filling filaments resemble those previously reported in amorphous Si \cite{Ugwumadu2025-SPTC-second-paper} and serve as the primary conduits for thermal transport in amorphous GST.

Evaluating its predictive performance at room temperature, we find an average thermal conductivity of 0.15  W/m$\cdot$K across the twelve amorphous GST models, in good agreement with the experimental value of $\approx$ 0.20  W/m$\cdot$K at room temperature \cite{Li2023-Thermal-Conductivity-aGST}. 

Two practical observations follow for considerations of materials design. First, the near-universal in-model ranking Te$\!\ge$Sb$\!\ge$Ge suggests that Te-rich (and, to a lesser extent, Sb-rich) regions form the dominant heat-carrying sub-network.  For example, in finite-element homogenization, weighting or refining around Te-dense clusters is therefore a sensible strategy when the quantity of interest is the effective conductivity \cite{Lewis1996,Tian2019}. Second, the “sensitivity’’ of M7 and M10 indicates that modest changes in medium-range order (e.g., a higher fraction of tetrahedral Ge, disrupted Te chains, or enhanced nanoscale porosity) can collectively depress all three site-projected channels, consistent with the sensitivity of amorphous chalcogenides to network topology and vibrational-mode localization.

\section{Conclusion}\label{sec:conclusion}

This study provides a novel characterization of amorphous Ge–Sb–Te structures by integrating electronic structure, vibrational analysis, and transport properties. Hybrid-functional calculations (HSE06) and the space-projected conductivity (SPC) reveal the role of mid-gap states, often associated with over-coordinated Ge atoms, in shaping electronic activity and conductivity. Thermally induced electronic fluctuations highlight the strong electron–phonon coupling at room temperature, especially for conduction tail states. 

On the thermal side, the site-projected thermal conductivity (SPTC) analysis reveals that Te- and Sb-rich regions form the dominant pathways for heat conduction (even taking into account the higher Te concentration due to the stoichiometry). Extended filamentary and chemically selective networks govern thermal transport in amorphous GST.

The combined use of SPC for electrons and SPTC for phonons establishes a framework for mapping charge and heat transport at the atomic scale. This spatially resolved perspective not only explains experimentally observed features, such as the low lattice thermal conductivity and the variability of electronic transport in GST, but also provides a generalizable methodology for analyzing transport in disordered materials. The SPTC statistics capture a chemically intuitive partitioning of heat flow, consistent with the broader literature on chalcogenides and amorphous vibrational transport.

\section*{Acknowledgement} 
We gratefully acknowledge Dr. D. Stewart for valuable discussions.

\section*{Funding} 
\noindent The computational resource used in this work was supported by the US National Science Foundation (NSF) under award number MRI2320493 and the Office of Naval Research under grant N000142312773. K. N.  acknowledges financial support from the Nanoscale \& Quantum Phenomena Institute (NQPI), conferred through the NQPI graduate research fellowship. K.K. acknowledges financial support from the Research Council of Finland under grant no. 364241 (“NoneqRSMSD”). S.R.E. is grateful to the Leverhulme Trust (UK) for a Fellowship. C. U. acknowledges funding from the Laboratory Directed Research and Development program of Los Alamos National Laboratory under the Director's Postdoctoral Fellowship Program, project number 20240877PRD4. Los Alamos National Laboratory is operated by Triad National Security, LLC, for the National Nuclear Security Administration of the U.S. Department of Energy (Contract No. 89233218CNA000001).

\bibliographystyle{elsarticle-num}
\bibliography{GST_main}

\begin{appendix}

\section{Spatially local estimates of transport}\label{app:appendixIntro}

For the atomistic local distribution of conductivity, we computed the space-projected conductivity (SPC), a spatial decomposition of the Kubo-Greenwood formula (KGF) for electronic conductivity, and site-projected thermal conductivity (SPTC)- a spatial decomposition of the Green-Kubo equation for thermal conductivity. The methodology for SPC and SPTC is detailed elsewhere \cite{SPC1, SPC2, Gautam2024-SPTC-first-paper, Ugwumadu2025-SPTC-second-paper}, but we provide a concise description here.

\subsection{Electronic Case: Space projected conductivity}\label{app:spc}
 Within linear response and the single particle approximation, the average of the diagonal elements of the conductivity tensor for each k-point, \textbf{k}, and frequency, $\omega$, is \cite{Kubo,Greenwood}:


\begin{align}
  \sigma_{\textbf k}(\omega) =& \frac{2\pi e^2}{3m^2\omega \Omega}\sum_{i,j} \sum_\alpha [f(\epsilon_{i,\textbf k})-f(\epsilon_{j,\textbf k})] \times \notag \\
 & {\mid \langle \psi_{j, \textbf k}|p^\alpha|\psi_{i,\textbf k} \rangle \mid}^2 \delta(\epsilon_{j,{ \textbf k}}-\epsilon_{i,{\textbf k}}-\hbar \omega)
\label{kgf_eqn}  
\end{align}

where \textit{e} and \textit{m} represent the charge and mass of the electron respectively. $\Omega$ represents the volume of the supercell. $\psi_{i,\textbf k}$ is the Kohn-Sham orbital associated with eigenvalue $\epsilon_{i,\textbf k}$. $f(\epsilon_{i,\textbf k})$ is the Fermi-Dirac weight and $p^\alpha$ is the momentum operator along Cartesian direction $\alpha$.

If we define $g_{ij}(\textbf {k},\omega)$ as \begin{equation}
g_{ij}({\textbf k},\omega)=\frac{2\pi e^2}{3m^2\omega \Omega} [f(\epsilon_{i,\textbf k})-f(\epsilon_{j,\textbf k})]\delta(\epsilon_{j,{\textbf k}}-\epsilon_{i,{\textbf k}}-\hbar \omega)
\end{equation}

The conductivity can then be expressed as (and dropping the \textbf{k} and $\omega$ dependence):

\begin{equation}
\sigma=\sum_{i,j,\alpha} g_{ij}\int d^3x\int d^3x^\prime[\psi_{j}^*(x)p^\alpha\psi_i(x)][\psi_{j}^*(x^\prime)p^\alpha\psi_i(x^\prime)]^*
\label{mom_oper_expansion}
\end{equation}

Taking $\xi_{ji}^\alpha(x)=\psi_j^*(x)p^\alpha\psi_i(x)$,  a complex-valued function defined on a real space grid (call the grid points \textbf{x}) with uniform grid size (\textit{h}) in three dimensions, then the integrals can be approximated as a sum on the grid from Eq.~(\ref{mom_oper_expansion}) yielding: 

\begin{equation}
\sigma \approx h^6\sum_{x,x^\prime}\sum_{i,j,\alpha} g_{ij}\xi_{ji}^\alpha(x)(\xi_{ji}^\alpha(x^\prime))^*
\label{sigma_approx}
\end{equation}

In Eq. (\ref{sigma_approx}), the approximation becomes exact as $h \rightarrow 0$.
Next, we define a Hermitian positive semi-definite matrix:

\begin{equation}
\Gamma(x,x^\prime) = \sum_{i,j,\alpha} g_{ij}\xi_{ji}^\alpha(x)[\xi_{ji}^\alpha(x^\prime)]^*
\label{gamma}
\end{equation}

from which:

\begin{equation}
\sigma = \sum_x \Gamma (x,x) + \sum_{x,x^\prime, x \ne x^\prime}\Gamma(x,x^\prime)
\label{gam}
\end{equation}

We show elsewhere that the object $\Gamma$ is interesting. Its eigenvalue problem reads $\Gamma \chi_\mu= \Lambda_\mu \chi_\mu$. $\Gamma$ always exhibits a huge null space, so that its eigenvectors conjugate to  a set of extremal $\Lambda$  offer a compact representation for the conduction active parts of the network.

The method has been implemented with VASP. It is possible to use a local orbital DFT Hamiltonian as in SIESTA \cite{SIESTA} or FIREBALL \cite{Lewis2011}, in which case one could extract site-projected SPC, which might be convenient for large systems.

\subsubsection{Electronic Conductivity and SPC in GST}\label{app:spc_method}

The wave-functions in equation \ref{kgf_eqn} are the Kohn-Sham orbitals $\Psi_{i;\textbf{k}}$ computed within VASP. The gradient of $\psi_{i}$ for each $\alpha$ was computed using the central finite difference method. The partial occupancies of the energy levels near the Fermi
level were approximated using the electronic temperature $T = 1000$ K for the Fermi-Dirac distribution. The delta function was approximated using a Gaussian of width 0.04 eV. To implement the SPC method, the 315-atom GST models were divided into a real space grid of 54 $\times$ 54 $\times$ 54  ($dim~\Gamma = 157464$). References \cite{Knyazev2013, Caldern2017, Bulanchuk2021, Subedi_2022} detail the dependence of KGF conductivity and SPC on different factors involving the KGF formalism.

\subsection{Electronic Case: $N^2 method$ \cite{N2}}\label{app:N2}

We next describe a simple and convenient method for both estimating the electrical conductivity of a material and extracting a spatial density function, which provides information about the spatial distribution of electrical conductivity. We have named this the $N^2$ method \cite{N2, W}, and tersely recount the derivation here. It follows from the work of Mott \cite{MottDavis, mott1969conduction} and Hindley \cite{RPA1, RPA2}, which indicates that electronic conductivity ($\sigma$) is determined by electronic activity near the Fermi energy ($\epsilon_f$). In particular, $\sigma$ is proportional to the square of the electronic density of states ($N(E)$) around $\epsilon_f$, expressed as \cite{N2}:

\begin{equation}\label{eqn:motts_sigma} 
\sigma \propto [N(E)]^2\big|_{E \rightarrow \epsilon_f} 
\end{equation}

This approximation is valid for systems with extended states near $\epsilon_f$. It is sometimes forgotten in the current literature that $N^2(\epsilon_f)$, \textit{not } $N(\epsilon_f)$ is a proxy for electronic conductivity \cite{deringer2021}.

The contribution of the $i^{th}$ Kohn-Sham orbital ($\epsilon_i$) to the electronic density of states at $\epsilon_f$ is:

\begin{equation}\label{eqn:N}
N(\epsilon_f) = \frac{1}{M} \sum_i \delta (\epsilon_i - \epsilon_f)
\end{equation}
where \textit{M} is the dimension of the single-particle Hamiltonian and in practice, the $\delta$ function is approximated by a Gaussian function of a selected smearing width. The spatial projection of the electronic conductivity can be expressed as \cite{N2}:

\begin{subequations}\label{eqn:N2}
\begin{gather}
\bar{\zeta}(\epsilon_f,\textbf{\textit{r}}) = \frac{1}{M^2} \sum_{i,j} \delta (\epsilon_i - \epsilon_f) \delta (\epsilon_j - \epsilon_f)|\psi_i(\textbf{\textit{r}})|^2|\psi_j(\textbf{\textit{r}})|^2\beta_{ij} \\
\beta_{ij} = \frac{1}{\int |\psi_i(\textbf{\textit{r}})|^2|\psi_j(\textbf{\textit{r}})|^2 d\textbf{\textit{r}}}.
\end{gather}
\end{subequations}

\noindent where $|\psi_i(\textbf{\textit{r}})|^2$ denotes the probability density of the $i^{th}$ Kohn-Sham orbital at the spatial grid point $\textbf{\textit{r}}$. The quantity $\beta_{ij}$ ensures that the volume integral of Equation \ref{eqn:N2} approximates $N^2(\epsilon_f)$ and $|\psi_i(\textbf{\textit{r}})|^2|\psi_j(\textbf{\textit{r}})|^2$ vanishes in grid regions where the wavefunctions do not overlap. Since each grid point $\textbf{\textit{r}}$ represents a contribution to $N^2$, their projection thus highlights electronic activity within the system.


\subsection{Thermal Case: site-projected thermal conductivity: SPTC}\label{app:sptc}

In a classic paper, Allen and Feldman (AF) worked out thermal transport from the Green-Kubo formula in the harmonic approximation, for disordered systems and a quantized lattice \cite{phonon1,phonon2}.  We have explained how to extract SPTC elsewhere \cite{Gautam2024-SPTC-first-paper, Ugwumadu2025-SPTC-second-paper} from their work, and for completeness, tersely repeat it here. At the $\Gamma$ point of the phonon Brillouin zone,  the vibrational normal modes are real.  For such a case,  the AF expression for TC is:

\begin{align} \label{eq:TC_Long}
    \kappa = & \frac{\pi \hbar}{48 T V} \sum_{m, n\neq m} \left[- \frac{\partial \langle f_m\rangle}{\partial \omega_m} \right] \delta(\omega_m - \omega_n ) \frac{\left( \omega_m + \omega_n \right)^2}{\omega_m \omega_n} \notag \\
    & \sum_{\eta} \sum_{\alpha, \beta} \sum_{\gamma, x, x'} e^{\alpha m}_{x} e^{\beta n}_{x'} \frac{1}{\sqrt{m_x m_{x'}}}  \phi^{\alpha \beta}_{x  x'} ({0}, \gamma) \left({R}^{\eta}_{\gamma} + {R}^{\eta}_{x x'}\right) \notag \\
    & \sum_{\alpha', \beta'} \sum_{\gamma', a, b} {e^{\alpha' m}_{a}} {e^{\beta' n}_{b}} \frac{1}{\sqrt{m_a m_b}}  \phi^{\alpha' \beta'}_{a b} ({0}, \gamma') \left({R}^{\eta}_{\gamma '} + {R}^{\eta}_{a b}\right)
\end{align}

where, $m$ and $n$ are the indices of the classical normal modes, $f_m$ is the equilibrium occupation of the $m^{th}$  mode, $\omega_m$ and $e^{\alpha m}_{i} $ are the vibrational frequency $\omega_m^\textbf{0}$ and the polarization $e^{\alpha m}_{i \textbf{0}}$ . ${R}_{\gamma}^{\eta}$ is the $\eta^{th}$ component of $\textbf{R}_{\gamma}$; ${R}_{x x'}^{\eta}$ is the $\eta^{th}$ component of $\textbf{R}_{x x'}$. Also, $\phi^{\alpha\beta}_{xx'}$ is the force constant tensor.  The thermal conductivity in Equation \ref{eq:TC_Long}, is taken as an average of the diagonal components of the conductivity tensor.

The AF form for TC may be rearranged as a double sum over spatial points (labeled ${x}$).   Carrying this out, with Equation \ref{eq:TC_Long}, we find:
\begin{equation}\label{sptc}
    \kappa =  \sum_{x, x'} \Xi(x, x'),
\end{equation}

\noindent where
\begin{align}
\label{eq:Gamma_def} 
   \Xi(x, x') &= \frac{\pi \hbar^2}{48k_BT^2V} \frac{1}{\sqrt{m_{x'} m_{x}}} \sum_{\eta} \sum_{\gamma}   \left({R}^{\eta}_{\gamma}  + {R}^{\eta}_{xx'}\right) \sum_{m, n \neq m} \delta(\omega_m - \omega_n) \notag \\
    &  \frac{(\omega_m + \omega_n)^2}{\omega_m \omega_n}  \left( \frac{e^{\frac{\hbar \omega_m}{k_B T}}}{\left(e^{\frac{\hbar \omega_m}{k_B T}} - 1\right)^2}\right)
    \sum_{\alpha \beta} \phi^{\alpha \beta}_{x, x'}(0, \gamma) e^{\alpha m}_{x} e^{\beta n}_{x'}  \notag \\
    &  \sum_{\gamma' a b} \sum_{\alpha' \beta'}\frac{1}{\sqrt{m_a m_b}}  \phi^{\alpha' \beta'}_{a, b} (0, \gamma')  e^{\alpha' m}_{a} e^{\beta' n}_{b} \left(R^{\eta}_{\gamma '} + R^{\eta}_{ab}\right).
\end{align}

$\Xi$ is called the "thermal matrix" which is a real-symmetric matrix with units of thermal conductivity. We decompose the total TC into contributions depending on atomic position $x$ by summing out one index of $\Xi(x, x')$ over positions $x '$:
\begin{equation}
\label{eq:zete_xi_relation}
\zeta(x) = \sum_{x'} \Xi(x, x').
\end{equation}
We call $ \zeta(x) $ the \textit{site-projected thermal conductivity} (SPTC),  the contribution of an atom at site $x$ in a supercell to the total TC of the system, since:
\begin{equation}
\label{eq:kappa_zeta}
    \kappa = \sum_{x} \zeta(x).
\end{equation}
For anisotropic systems or off-diagonal terms, the conductivity tensor $\kappa_{\alpha \beta}$ local contributions can be similarly obtained.
 
A simple argument clarifies the physical meaning of SPTC. The decomposition above expresses the thermal conductivity as a sum over all pairs of sites. Now, imagine removing one atom from the cell at position $y$. The difference in thermal conductivity between the original system and the one with the atom missing at $y$ is evidently (by excluding all the terms involving the site $y$ in the SPTC) $\Delta_y=2\sum_{x'}\Xi({x',y)}$  which we can call the "Atomic Removal Conductivity" (ARC) for site $y$ and we used the fact that the thermal matrix, $\Xi$, is symmetric.  The term $\sum_{x'}\Xi(x',y)$ is exactly the SPTC for site $y$ obtained from the preceding discussion, so that $\zeta_\kappa{(y)}=\Delta_y/2 $.   \textit{We conclude that the SPTC at site $y$ is just half of the ARC}. 

Electrons and phonons both contribute to the thermal conductivity in GST, but we will limit ourselves to the lattice contribution (phonons). We will be implementing SPTC as a tool for analyzing the thermal conductivity, which is based on the Green-Kubo formula. A cutoff distance of $r_c = 8~ $ \AA{} was used for the calculation based on decay analysis described in Figure \textcolor{blue}{S7}.
\end{appendix}

\end{document}


\title{SUPPLEMENTARY MATERIAL\\Electronic and thermal properties of the phase-change memory material, Ge$_2$Sb$_2$Te$_5$, and results from spatially resolved transport calculations.}

\author[OU]{K. Nepal}
\ead{kn478619@ohio.edu}
\affiliation[OU]{organization={Department of Physics and Astronomy, Nanoscale and Quantum Phenomena Institute (NQPI)},
            addressline={Ohio University}, 
            city={Athens},
            postcode={45701}, 
            state={OH},
            country={USA}}

\author[OU]{A. Gautam}

\author[OU]{R. Hussein}

\author[Tampere]{K. Konstantinou}
\affiliation[Tampere]{organization={Department of Mechanical and Materials Engineering},
            addressline={University of Turku}, 
            city={Turku},
            postcode={FI-20500}, 
            country={Finland}}
            

\author[SRE]{S. R. Elliott}
\affiliation[SRE]{organization={Physical and Theoretical Chemistry Laboratory, Department of Chemistry},
            addressline={University of Oxford}, 
            city={Oxford},
            postcode={OX1 3QZ}, 
            country={UK}}

\author[LANL]{C. Ugwumadu}
\ead{cugwumadu@lanl.gov}
\affiliation[LANL]{organization = Physics of Condensed Matter and Complex Systems (T-4) Group, addressline = { Los Alamos National Laboratory}, city = { Los Alamos}, postcode = {87545}, state ={NM}, country={USA}}

\author[OU]{D. A. Drabold}
\ead{drabold@ohio.edu}

\date{July 2025}

\begin{abstract}
We report new insights into the electronic, structural, and transport (heat and charge) properties of the phase-change memory material Ge$_2$Sb$_2$Te$_5$. Using realistic structural models of Konstantinou \textit{et al.} [Nat. Commun. 10, 3065 (2019)], we analyze the topology, electronic states, and lattice dynamics with density functional methods, including hybrid-functional calculations and machine-learned interatomic potentials. The Kohn–Sham orbitals near the Fermi level display a strong electron–phonon coupling, and exhibit large energy fluctuations at room temperature. The conduction tail states exhibit larger phonon-induced fluctuations than the valence tail states. To resolve transport at the atomic scale, we employ space-projected electronic conductivity and site-projected thermal conductivity methods.  Local analysis of heat transport highlights the role of filamentary networks dominated by Te, with Sb and Ge making progressively smaller contributions.
\end{abstract}
\maketitle

\section{Tables}
Table \ref{tab:MAD} shows the deviation from rock-salt 3-body angle of 90$^\circ$ in the 12 GST models. It should be pointed out that in the ideal rock salt crystal structure (e.g., sodium chloride, NaCl), all 3-body angles between nearest-neighbor ions are 90° and 180°. The metastable crystalline phase of the phase-change material Ge-Sb-Te (GST) has a rock-salt structure with significant disorder \cite{Sun2006,Cinkaya2011}, and is crucial for its role in phase-change memory (PCM) applications. In this structure, tellurium (Te) occupies one sublattice, while germanium (Ge), antimony (Sb), and intrinsic vacancies occupy the other sublattice, forming a disordered but layered structure that enables the material's function in data storage. Deviations occur in the amorphous phase, as seen in Table \ref{tab:MAD}.

\begin{table}[!t]
\centering
\scriptsize
\caption{{Mean absolute deviation (MAD) from rock-salt 3-body angle of 90$^\circ$ in the 12 GST models. }}
\footnotesize
\begin{tabularx}{\linewidth}{@{\extracolsep\fill}cccc@{}}
\toprule
\textbf{Model} & \textbf{Ge} & \textbf{Sb} & \textbf{Te} \\
\midrule
M1 & 12.24 & 11.98 & 12.77 \\ 
M2 & 12.16 & 11.66 & 12.99  \\ 
M3 & 12.41 & 11.15 & 13.04  \\ 
M4 & 12.44 & 11.74 & 12.76  \\ 
M5 & 12.61 & 11.41 & 12.87  \\ 
M6 & 12.65 & 11.41 & 13.07  \\ 
M7 & 12.72 & 11.86 & 12.99  \\ 
M8 & 12.65 & 11.77 & 12.71  \\ 
M9 & 12.37 & 10.94 & 12.74  \\ 
M10 & 12.02 & 11.96 & 12.89  \\ 
M11 & 12.16 & 11.57 & 12.90  \\ 
M12 & 12.70 & 11.29 & 12.95 \\
\bottomrule
\end{tabularx}
\label{tab:MAD}
\end{table}

\newpage
\section{Figures}

\begin{figure}[!h]
    \centering
	\includegraphics[width=\linewidth]{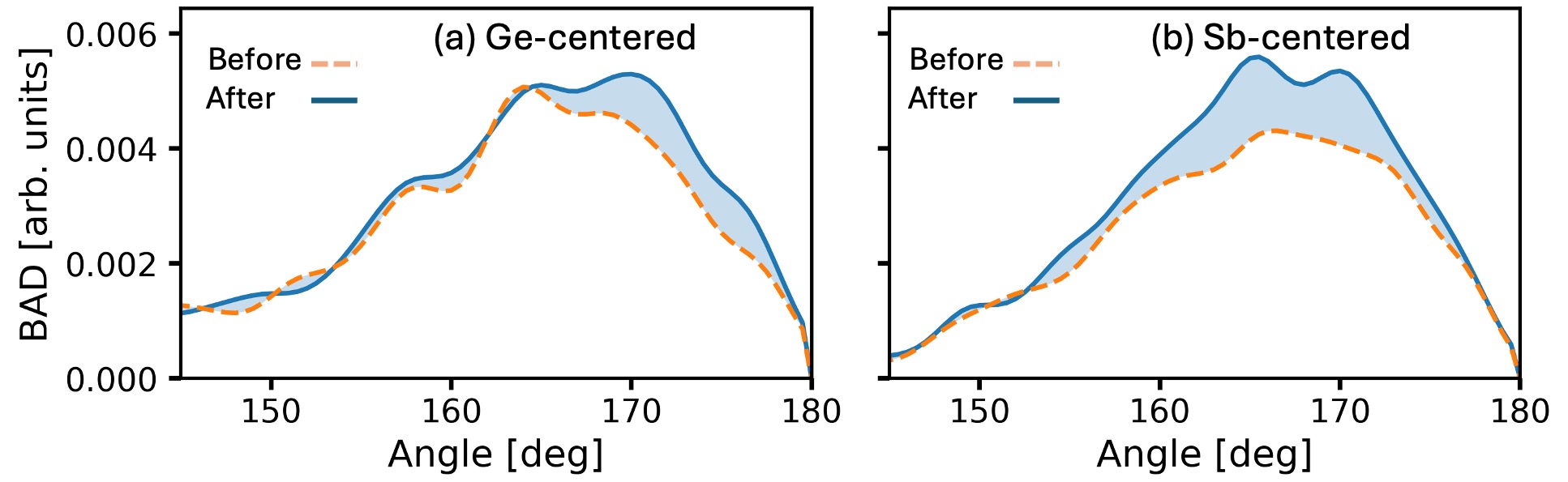}
    \caption{ Bond angle distribution (BAD) for (a) Ge-centered and (b) Sb-centered local environment averaged over twelve GST models from KMLE after HSE06 relaxation (Blue color). BAD was calculated by using a cut-off bond distance of 3.2 \AA{} for each atom species. The dashed orange plot shows the average BAD calculated from before HSE06 relaxation.}
    \label{fig:Sfig_bad}
\end{figure}

\begin{figure*}[!h]
    \centering
	\includegraphics[width=\linewidth]{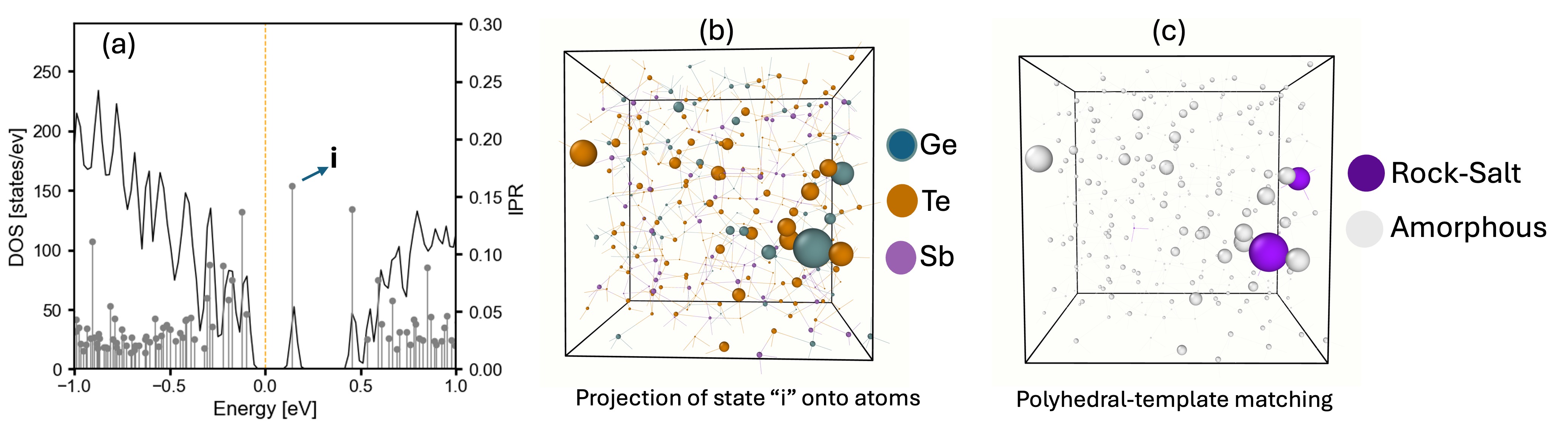}
    \caption{Electronic properties of mid-gap states in GST model M5. Electronic density of states and inverse participation ratio, depicting the mid-gap state “i”. The yellow dashed line represents the Fermi level (shifted to 0). The projection of the mid-gap state into atomic species (see legends for atomic species), shown in (b), shows that the state is localized predominantly on an overcoordinated Ge atom (larger sphere $\implies$ higher contribution). However, about 6 Te atoms contribute considerably to this state. We performed a polyhedral template matching analysis \cite{Larsen2016-Polyhedral-Template-Matching} in (c), which indicates that the most localized Ge atom has a rock-salt-like (crystal) local bonding environment.}
    \label{fig:Sfig_midGap_defect}
\end{figure*}

\begin{figure*}[!h]
    \centering
	\includegraphics[width=0.8\textwidth]{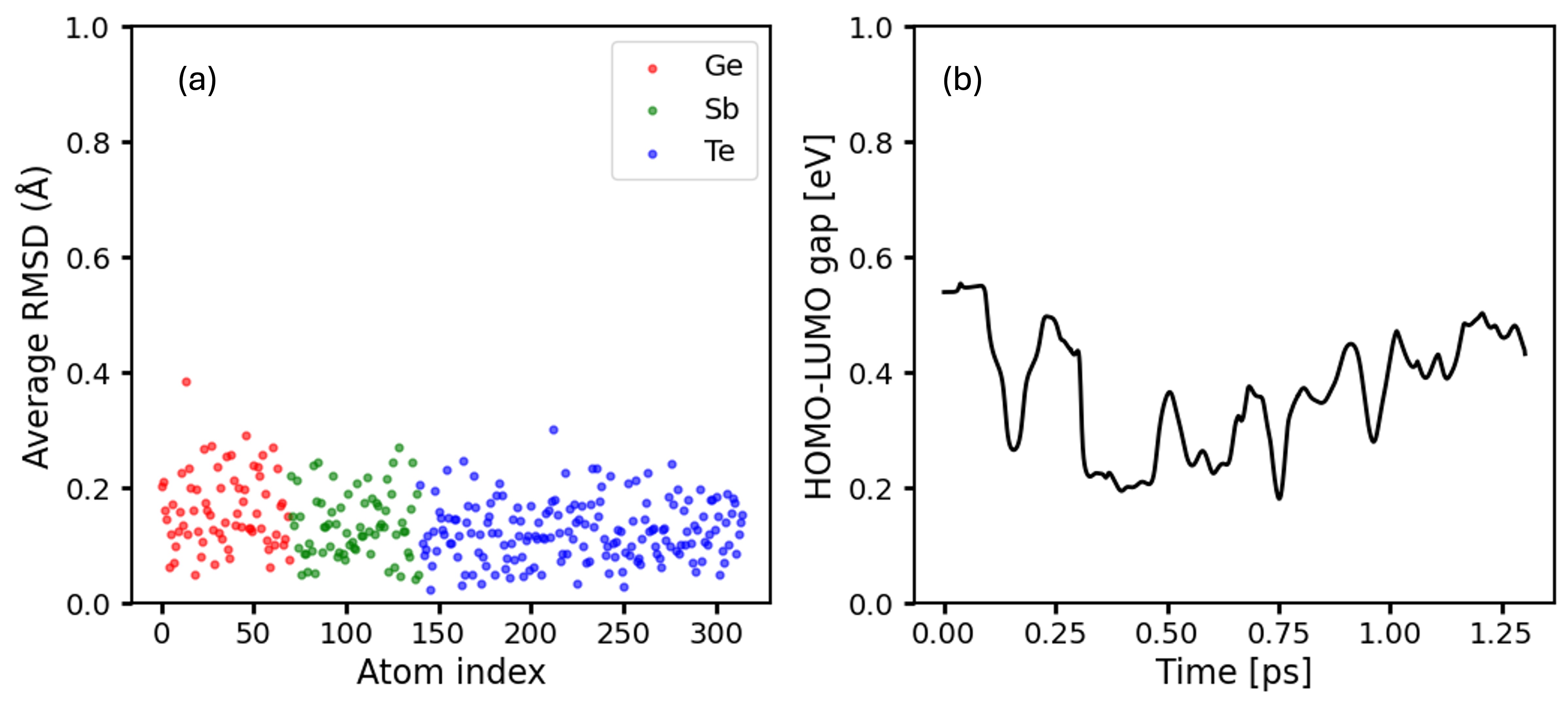}
    \caption{ Average root mean square deviation (RMSD) of atomic species over the last 150 fs, and (b) fluctuation of the HOMO–LUMO gap during molecular dynamics at 300 K for model M5, equilibrated at 300 K using a Nosé–Hoover thermostat with a time step of 1.5 fs. }
     \label{fig:Sfig_rmsd_gap}
\end{figure*}

\begin{figure*}[!h]
    \centering
	\includegraphics[width=\linewidth]{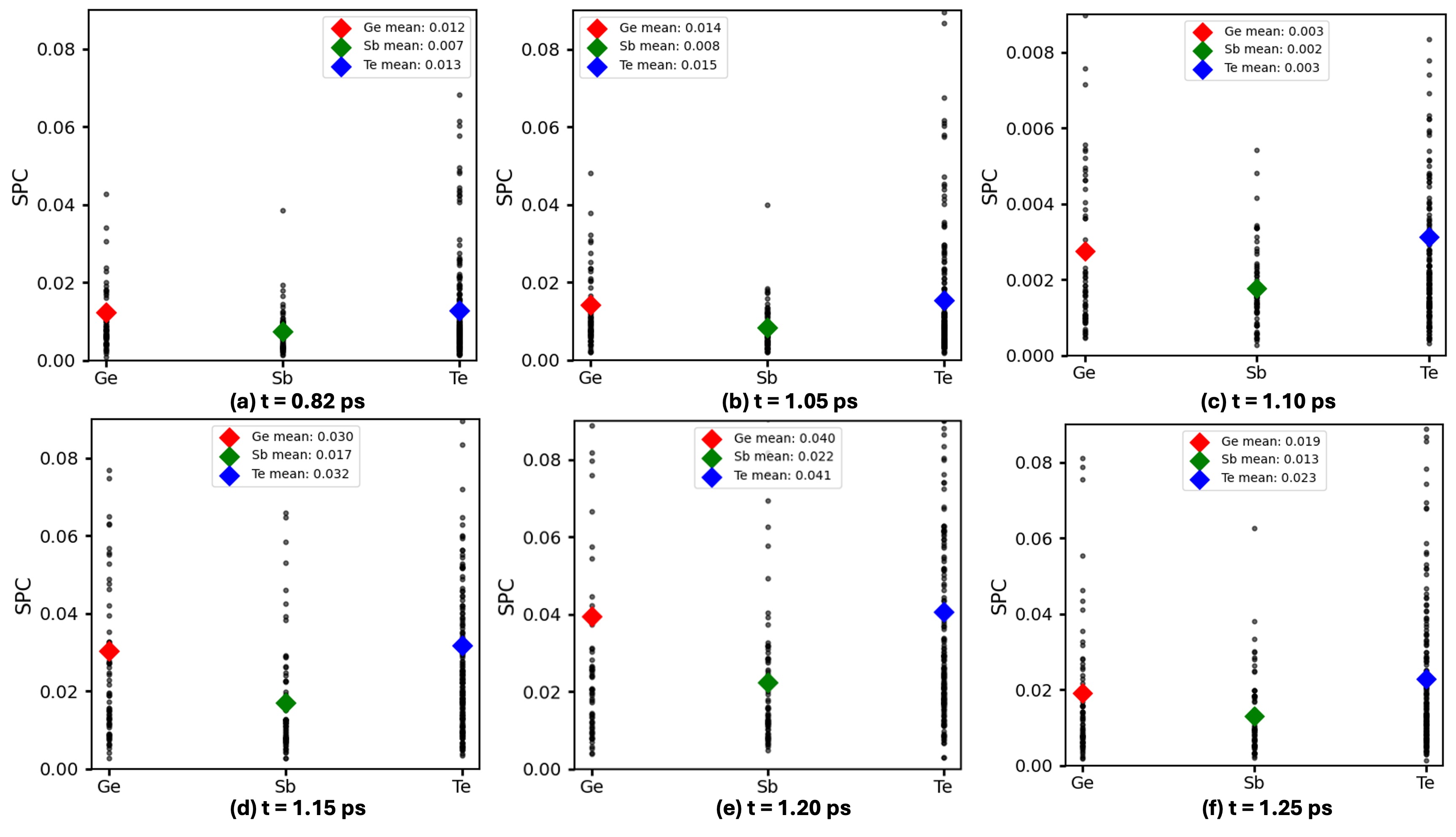}
    \caption{Atomic species contribution to the space-projected conductivity (SPC in units Siemens/cm) of snapshots during molecular dynamics at 300 K for model M5, equilibrated at 300 K using a Nosé–Hoover thermostat with a time step of 1.5 fs. To compute the atomic species contribution, we interpolate SPC from the real-space grid points to the atomic species.  The SPC values are scaled by the factor of 10$^4$ in Figures (a-c) and (d), while scaled by a factor of 10$^2$ in Figures (d-e).}
    \label{fig:Sfig_site_SPC}
\end{figure*}

\begin{figure}[!h]
    \centering
    \includegraphics[width=1.0\linewidth]{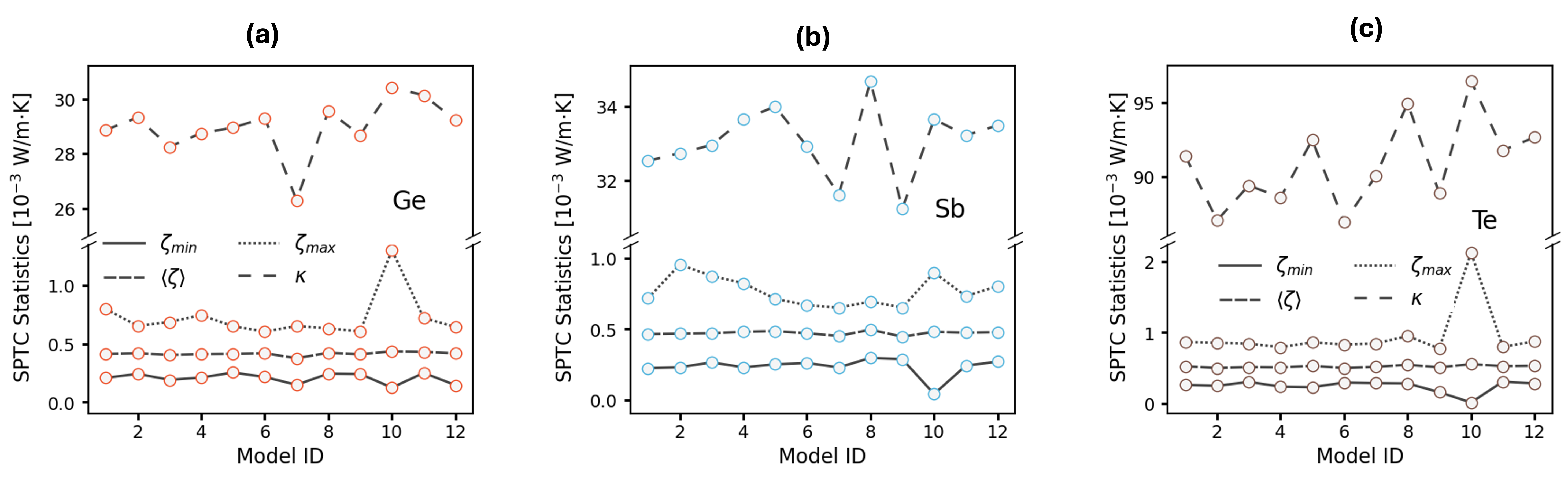}
    \caption{Thermal transport character in GST. (a -- c) element-specific SPTC statistics for models M1 to M12. }
    \label{fig:Sfig_GST_SPTC_Decay}
\end{figure}

\begin{figure}[!t]
    \centering
    \includegraphics[width=\linewidth]{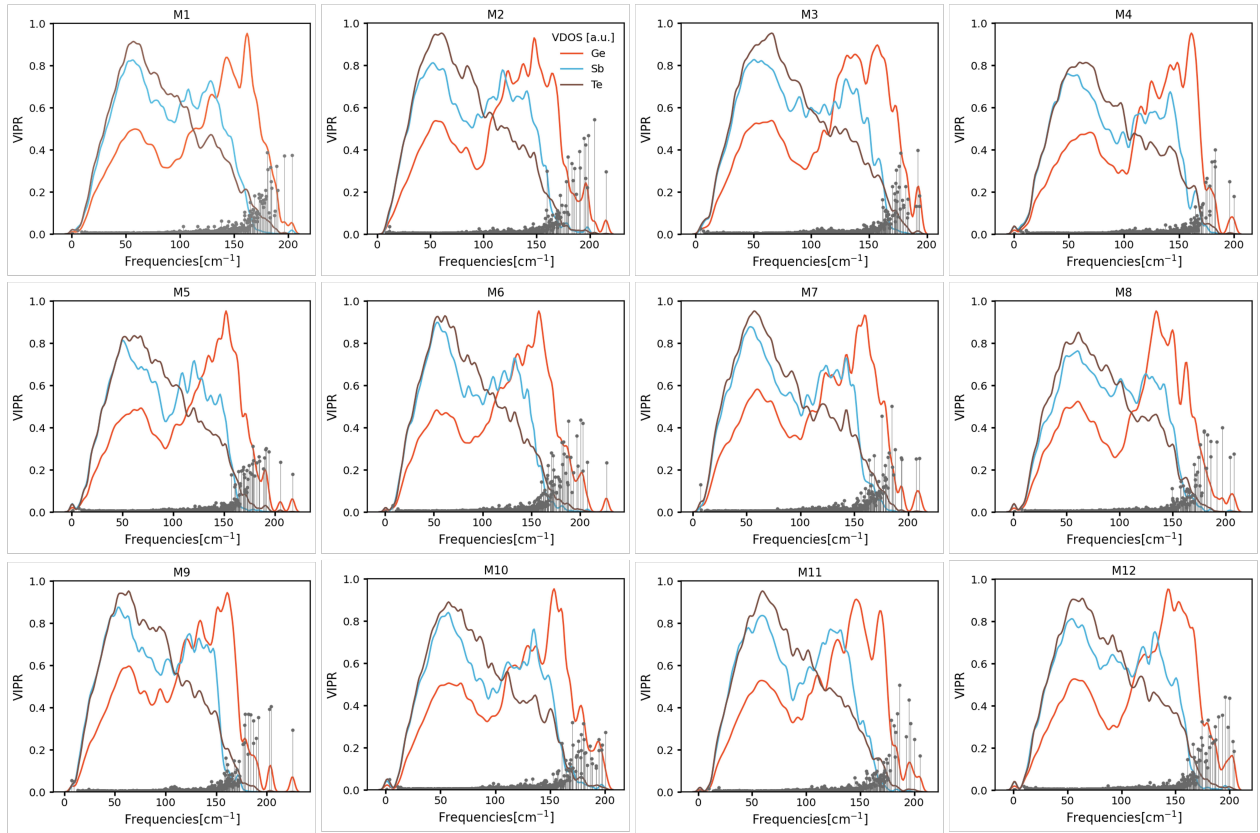}
    \caption{Partial vibrational density of states (VDOS) and vibrational inverse participation ratio (VIPR) for Models M1 -- M12. As discussed in the main text, Te contributes mostly to the low-mid frequency regime, the diffusons regime, which significantly influences SPTC distribution, and hence thermal transport. The highly localized Ge modes in the high-frequency regime (locons) contribute minimally to thermal transport. Ge is highly localized at the high-frequency regime (locons), which minimally contributes to thermal transport. Sb shows peaks in the low and mid-frequency range and also shows moderate SPTC contribution in the systems. The VDOS are all comparable to those in Figure 16 of reference \cite{Caravati2007}.}
    \label{fig:Sfig_PVDOS}
\end{figure}

\begin{figure}[!t]
    \centering
    \includegraphics[width=\linewidth]{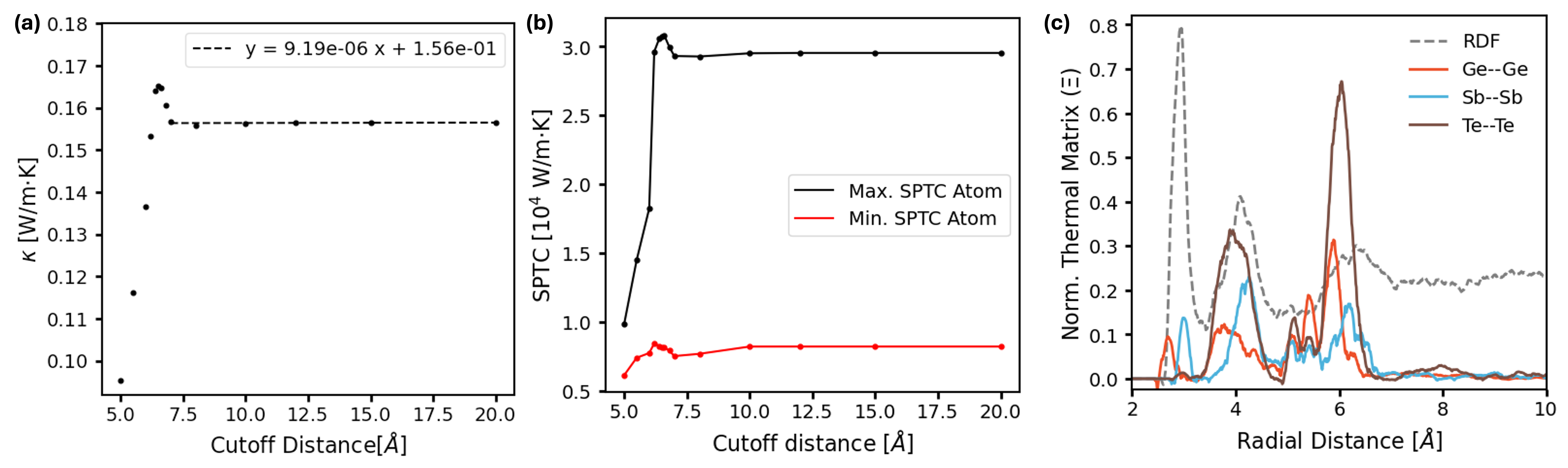}
    \caption{Total thermal conductivity ($\kappa$) of a 900-atom amorphous GST model with increasing cutoff distance is given in Figure (a). The value of the conductivity is relatively constant beyond a cutoff distance of $r_c = 7$\AA{}. This is supported by the convergence of SPTC values of minimum and the maximum SPTC atoms in the system, given in Figure (b). (c) The partial (per-element) thermal matrix ($\Xi(x, x')$) distribution for a 900-atom amorphous GST structure, plotted alongside its radial distribution function (RDF).}
    \label{fig:Sfig_k_convergence}
\end{figure}

\begin{figure*}[!t]
    \centering
    \includegraphics[width=\linewidth]{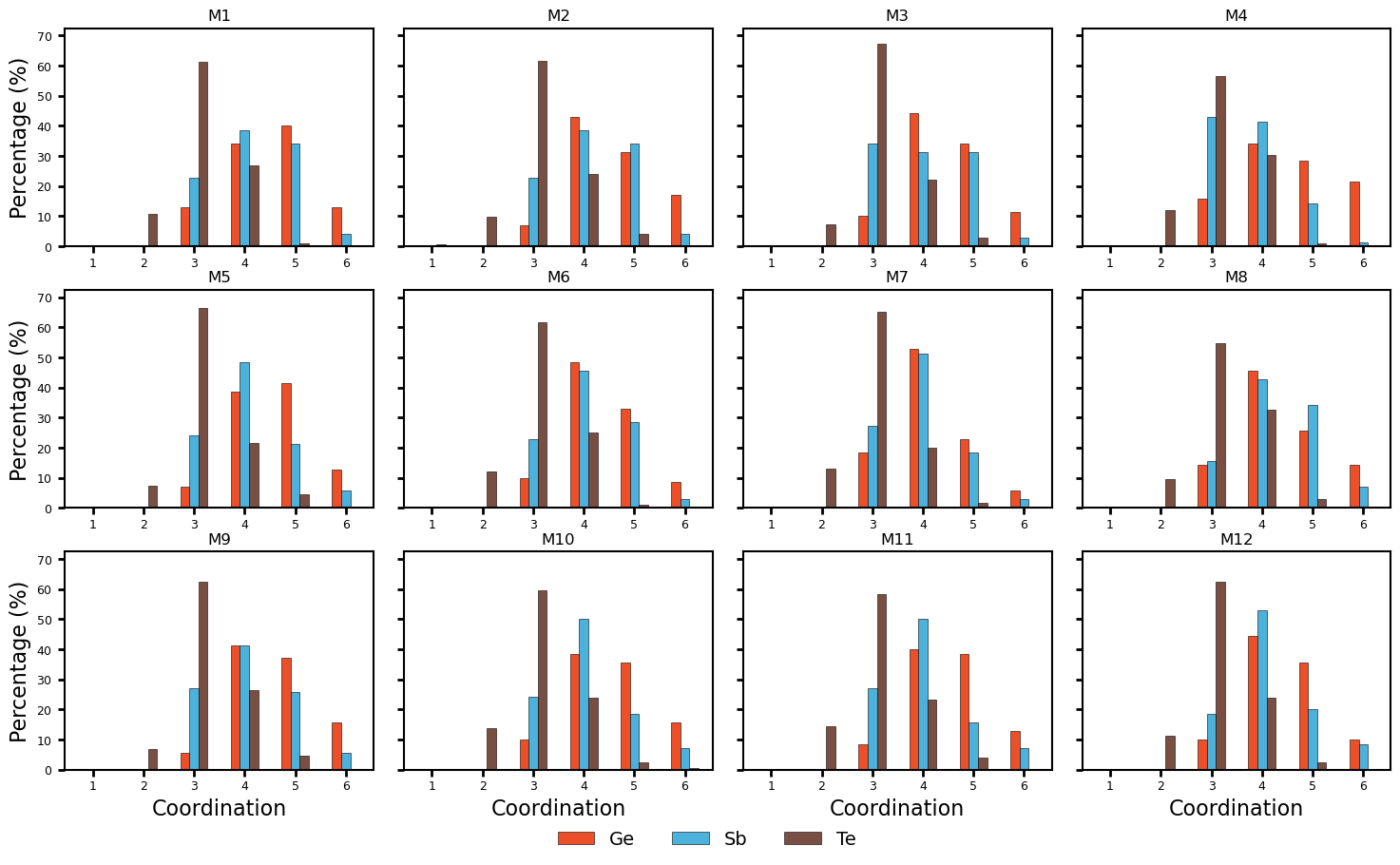}
    \caption{Element-specific CN distribution for all
    Ge, Sb, Te configurations (models M1-M12). In each panel, there is consistency for Te atoms, mainly dominating three-fold coordination, whereas Sb exhibits three and four-fold coordination. Ge coordination is mostly dominated by 4-coordinated tetrahedral and defective octahedral sites. While Ge mostly dominates the tetrahedral sites, with a few present in the fivefold sites. This consistency underscores the strength of the local bonding structure.}
    \label{fig:Sfig_CN}
\end{figure*}

\newpage

\bibliographystyle{elsarticle-num}
\bibliography{GST_main}%